\documentclass[iop]{emulateapj}
\slugcomment{{\sc Accepted for publication in AJ}}
\usepackage[section]{placeins}

\def\deg{\hbox{$^{\:\circ}$}}

\slugcomment{}

\shorttitle{Nuclear IR Emission of LLAGN}
\shortauthors{Mason et al.}

\begin{document}

\title{The Nuclear Infrared Emission of Low-Luminosity Active Galactic Nuclei}

\author{R. E. Mason}
\affil{Gemini Observatory, Northern Operations Center, 670 N. A'ohoku Place, Hilo, HI 96720, USA}

\author{E. Lopez-Rodriguez, C. Packham}
\affil{University of Florida, Department of Astronomy, 211 Bryant Space Science Center, P.O. Box 112055, Gainesville, FL 32611, USA}

\author{A. Alonso-Herrero}
\affil{Instituto de F\'{\i}sica de Cantabria, CSIC-UC, Avenida de los Castros
s/n, 39005 Santander, Spain}
\affil{Augusto Gonz\'alez Linares Senior Research Fellow}

\author{N. A. Levenson, J. Radomski}
\affil{Gemini Observatory, Southern Operations Center, c/o AURA, Casilla 603, La Serena, Chile}

\author{C. Ramos Almeida}
\affil{Instituto de Astrof\'\i sica de Canarias, 
              C/V\'\i a L\'{a}ctea, s/n, E-38205, La Laguna, Tenerife, Spain; Departamento de Astrof\' isica, Universidad de La Laguna,   
              E-38205, La Laguna, Tenerife, Spain.}

\author{L. Colina}
\affil{Departamento de Astrof\'{\i}sica,
Centro de Astrobiolog\'{\i}a (CSIC/INTA),
Instituto Nacional de T\'{e}cnica Aeroespacial,
Crta de Torrej\'{o}n a Ajalvir, km 4,
28850 Torrej\'{o}n de Ardoz, Madrid
Spain}

\author{M. Elitzur}
\affil{Department of Physics and Astronomy, University of Kentucky, Lexington, KY 40506, USA}

\author{I. Aretxaga}
\affil{Instituto Nacional de Astrof\'{i}sica, \'{O}ptica y Electr\'{o}nica (INAOE), Aptdo. Postal 51 y 216, 72000 Puebla, Mexico}

\author{P. F. Roche}
\affil{Astrophysics, Department of Physics, University of Oxford, DWB, Keble Road, Oxford OX1 3RH, UK}

\author{N. Oi}
\affil{Department of Astronomy, School of Science, Graduate University for Advanced Studies (SOKENDAI), Mitaka, Tokyo 181-8588, Japan}

\begin{abstract}

We present high-resolution mid-infrared (MIR) imaging, nuclear spectral energy distributions (SEDs) and archival Spitzer spectra for 22 low-luminosity active galactic nuclei (LLAGN; L$_{\rm bol} \lesssim 10^{42} \; \rm erg \; s^{-1}$).   Infrared (IR) observations may advance our understanding of the accretion flows in LLAGN, the fate of the obscuring torus at low accretion rates, and, perhaps, the star formation histories of these objects. However, while comprehensively studied in higher-luminosity Seyferts and quasars, the nuclear IR properties of LLAGN have not yet been well-determined. We separate the present LLAGN sample into three categories depending on their Eddington ratio and radio emission, finding different IR characteristics for each class. (I) At the low-luminosity, low-Eddington ratio (log $\rm L_{bol} / L_{Edd} < -4.6$) end of the sample, we identify "host-dominated" galaxies with strong polycyclic aromatic hydrocarbon bands that may indicate active (circum-)nuclear star formation. (II) Some very radio-loud objects are also present at these low Eddington ratios. The IR emission in these nuclei is dominated by synchrotron radiation, and some are likely to be unobscured type 2 AGN that genuinely lack a broad line region. (III) At higher Eddington ratios, strong, compact nuclear sources are visible in the MIR images. The nuclear SEDs of these galaxies are diverse; some resemble typical Seyfert nuclei, while others lack a well-defined MIR ``dust bump''. Strong silicate emission is present in many of these objects. We speculate that this, together with high ratios of silicate strength to hydrogen column density, could suggest optically thin dust and low dust-to-gas ratios, in accordance with model predictions that LLAGN do not host a Seyfert-like obscuring torus. We anticipate that detailed modelling of the new data and SEDs in terms of accretion disk, jet, radiatively inefficient accretion flow and torus components will provide further insights into the nuclear structures and processes of LLAGN.

\end{abstract}


\section{Introduction}
\label{intro}

An active galaxy spends only a small fraction of its lifetime in a spectacular, highly luminous, quasar-like phase \citep[e.g.][]{Hopkins06,Shankar09}. Much more time is spent in a near-quiescent, weakly-accreting state, and indeed low-luminosity AGN (LLAGN)  are found in nearly half of nearby galaxies \citep{Ho08}. The properties of the accretion flow and associated structures in LLAGN are thought to be quite different from those in their higher-luminosity counterparts. A full description of AGN accretion and lifecycles, then, will require understanding the low-luminosity members of the class.

In this paper we investigate the nuclear infrared (IR) properties of LLAGN\footnotemark. The IR continuum emission from an AGN may contain contributions from a variety of processes: thermal emission from a dusty torus surrounding the central supermassive black hole (SMBH), synchrotron emission from a jet, and -- in LLAGN -- thermal emission from a truncated accretion disk (see below). Dust in a narrow-line region (NLR) or associated with the surrounding stellar population may also contribute. Silicate emission or absorption features at 10 and 18 $\mu$m contain information about dust geometry and heating, and the suite of IR polycyclic aromatic hydrocarbon (PAH) bands may trace star formation in the nuclear environment. The IR wavelength regime therefore contains much information that could potentially advance our understanding of the nature and lifecycle of LLAGN. However, IR observations of LLAGN remain fairly scarce, especially at the angular resolution necessary to isolate the weak nuclear emission from that of the host galaxy. 

\footnotetext{For the purposes of this paper, we loosely define a ``low-luminosity'' AGN as a low-ionisation nuclear emission region (LINER) or Seyfert galaxy with $L_{\rm bol}$ below about $10^{42}\; \rm erg\; s^{-1}$. For comparison, Sgr~A* has $L_{\rm bol} \sim 10^{36} \; \rm erg\; s^{-1}$, the dwarf Seyfert NGC~4395 has $L_{\rm bol} \sim 10^{40} \; \rm erg\; s^{-1}$ and the luminous quasar PDS~456, $L_{\rm bol} \sim 10^{47} \; \rm erg\; s^{-1}$.}

Studies of the spectral properties, luminosities, fuel supply etc. of LLAGN indicate that the SMBHs in these objects are accreting at low rates and with low radiative efficiency \citep{Yuan07,Ho09,Trump11,Yu11}. These studies suggest that the standard optically thick and geometrically thin accretion disk is truncated in LLAGN, and replaced by a geometrically thick, radiatively inefficient accretion flow \citep[RIAF; e.g.][]{Rees82,Narayan94,Narayan08} interior to the disk truncation radius. In addition, jet emission becomes increasingly important as the accretion rate decreases \citep{Yuan05}. The truncation of the thin disk should shift its thermal emission peak towards longer wavelengths, leading to the suggestion that the mid-IR (MIR) peaks or excesses observed in the SEDs of some LLAGN are the equivalent of the ``big blue bump'' of Seyferts and quasars \citep{Ho08}. Indeed, \citet{Nemmen11} present models in which the bulk of the nuclear luminosity at $\sim$1--10 $\mu$m can come from a truncated thin disk. However, the authors note that the available, large-aperture IR data are not sufficient to constrain the presence or properties of a truncated disk in the objects modeled. IR photometry, at higher angular resolution than is generally available in the literature,  may therefore aid our understanding of the accretion physics of LLAGN. 

In higher-luminosity AGN, the difference between types 1 and 2 is explained at least to first order by the presence of a toroidal cloud of dust and gas obscuring the AGN from certain viewing directions while permitting a direct view from others \citep{Antonucci93}. An obvious question, then, is how (if at all) do LLAGN fit into this unified framework?

Some models of the torus explain its existence through inflows of gas from larger scales, often invoking the effects of nuclear star clusters or disks \citep[e.g.][]{Wada09,Schartmann10,Hopkins11}. During periods when little material is reaching the centre of the galaxy, the torus may become thin and transparent \citep{Vollmer08}. Conversely, the torus may be part of a dusty, outflowing wind \citep{Konigl94,Elitzur06,Dorodnitsyn11}. The disk wind model of \citet{Elitzur06} predicts that below L$_{\rm bol} \sim 10^{42}\; \rm erg \; s^{-1} $, accretion onto the black hole can no longer sustain the outflow necessary to obscure the nucleus. In either case, low-luminosity AGN may show little nuclear obscuration and dust emission. In the \citet{Elitzur06} model, very low-luminosity AGN are also expected to lack a broad line region \citep[see also][]{Nicastro00,Laor03,Elitzur09}. 

Observationally, there are indications that LLAGN do tend to have unobscured nuclei. For instance, several authors have detected nuclear UV/optical point sources in LINERs of both types 1 and 2 \citep{Pogge00,Chiaberge05,Maoz05}. In a study of  86 nearby galaxies with pointlike X-ray nuclei and L$_{\rm X} < 10^{42}\; \rm erg \; s^{-1}$, \citet{Zhang09} find $N_{\rm H}$ to be correlated with X-ray luminosity. \citet{Brightman11} also find a decline in obscuration at low luminosities. Detailed X-ray observations of some LLAGN have shown that the Fe K$\alpha$ line, a signature of X-ray reprocessing by cool material, is weak in these nuclei  \citep{Ptak04,Binder09}.

However, this is not the case for all objects. The detection of broad H$\alpha$ in polarised light in 3/14 LINERs \citep{Barth99} suggests that these objects host dust-obscured AGN, as do the biconical ionization cones observed in a handful of LINERs \citep{Pogge00}. X-ray studies of single objects show that some LLAGN do have substantial absorbing columns \citep[e.g. NGC~4261;][]{Zezas05}, and the fraction of such objects may be significant \citep{Gonzalez-Martin09b}. \citet{Sturm06} present average spectra of type 1 and 2 LINERs which suggest an extra hot dust component present in the type 1s relative to the type 2s, similar to the SEDs of ``conventional'' Seyferts \citep{Alonso-Herrero03,Deo09,Prieto10,RamosAlmeida11}.  
The role of dust in LLAGN, and in particular whether there exists a luminosity, accretion rate, or other property at which the torus ceases to exist, remains unclear. A search for the IR signatures of the torus -- thermal emission and silicate emission and absorption features -- promises a better understanding of these issues. 

Finally, IR observations may give information about the star formation histories of LLAGN.  Based on optical spectroscopy, LINER stellar populations are generally thought to be old \citep{Ho03,Kewley06,Zhang08}, although intermediate-age populations have been detected in a minority of cases \citep{GonzalezDelgado04}. However, the optical regime is not ideally suited for discovering very young stars ($\sim$ few Myr) or star formation activity. Small-aperture UV spectroscopy of stellar absorption lines can be a useful technique \citep{Maoz98,StorchiBergmann05}, but is vulnerable to extinction. Although not always an unambiguous star formation indicator \citep{Peeters04,Diaz-Santos10}, observations of PAH emission in LLAGN could prove to be a complementary method of tracing young stellar populations and/or ongoing star formation \citep[e.g.][]{Shapiro10}. IR spectroscopy can also probe older stellar populations, which may be important in explaining the energy budgets of LLAGN \citep{Eracleous10b}, by locating spatially extended silicate emission features from AGB star envelopes \citep{Bressan06,Buson09}.

At the time of writing, the IR emission of LLAGN remains comparatively unexplored, particularly on spatial scales $<$ 1\arcsec.  For many years following early, lower-resolution MIR work \citep[e.g.][]{Roche85, Lawrence85,Willner85, Cizdziel85, Roche91,Knapp92},  high-resolution imaging at $\lambda \sim 10 \; \mu$m existed only for a handful of relatively bright, well-known objects \citep{Chary00,Grossan01,Perlman01,Whysong04,Mason07,Horst08, Radomski08,Reunanen10}. More recently, \citet{Asmus11} have presented ground-based MIR imaging of a number of LLAGN, with seven new detections. In addition, \citet{vanderWolk10} detected 10 radio galaxies in their high-resolution MIR images, some of which can be considered LLAGN. Some subarcsecond-resolution NIR imaging is also available in the literature, but it has rarely been considered in the context of the multi-wavelength emission of LLAGN. Published LLAGN SED compilations contain little or no high-resolution IR data \citep{Ho99,Ptak04,Maoz07,Eracleous10a}.

With the aim of illustrating the overall nuclear IR properties of LLAGN, we have acquired new, ground-based MIR ``snapshot'' imaging of 20  IR-faint nuclei. Acquired with Michelle and T-ReCS on the Gemini telescopes, the angular resolution of the observations is approximately 0\arcsec.35 at $\lambda \sim$10~$\mu$m. We have also obtained seeing-limited images at 3 -- 5 $\mu$m of five objects. To the new data, we add published, high-resolution IR photometry of these and a further two nuclei. We combine these data with published measurements at other wavelengths to produce nuclear spectral energy distributions (SEDs) for these objects. In terms of high spatial resolution IR data, these SEDs are by far the most detailed yet available. Finally, we also present archival Spitzer low-resolution spectroscopy for the 18/22 galaxies with available data. 

In \S\ref{sample} we discuss the galaxy sample studied in this paper. \S\ref{obsdr} describes the reduction of the MIR and NIR observations and the resulting photometric measurements. In \S\ref{img} we present the images and use the MIR/X-ray relation to investigate the nature of the structures revealed in them. \S\ref{irs} shows the IRS spectra of the LLAGN, and the full, high-resolution radio -- X-ray SEDs are presented in \S\ref{sect:seds} (with a detailed view of the 1 -- 20 $\mu$m region in Appendix \ref{app1}). The processes that may give rise to the IR emission in the different types of LLAGN are discussed in \S\ref{discuss}. We anticipate that the new IR data and SEDs will provide valuable constraints on models of the nuclear processes in LLAGN, but we defer such detailed modelling to future work.

\section{The LLAGN Sample}
\label{sample}

The 22 LLAGN studied in this paper have L$_{2-10 \rm \; keV} < 3 \times 10^{41} \rm \; erg \; s^{-1}$ (corrected for extinction; see \S\ref{sect:seds}) and D $<$ 36 Mpc (1\arcsec $<$ 170 pc). Assuming L$_{\rm bol} / \rm L_{X} \sim 16$ \citep{Ho09}\footnotemark, this X-ray luminosity corresponds to L$_{\rm bol} \sim 5 \times 10^{42} \; \rm erg \; s^{-1}$, roughly the threshold below which \citet{Elitzur06} predict the disappearance of the torus in LLAGN.  Spectral classifications were taken from \citet{Ho97}. Most of the objects are LINERs, but the sample also contains a few low-luminosity Seyfert galaxies. Basic information for all of the sources is given in Table \ref{basic_data}. 

\footnotetext{All bolometric luminosities (and, therefore, Eddington ratios) given in this paper are derived using this conversion factor unless otherwise stated. While a convenient tool, X-ray-based bolometric corrections are still a matter of some debate; see, for example, \citet{Eracleous10a}.}

All of the galaxies we observed had old, low-resolution IR photometry available in the literature \citep[e.g.][]{Willner85,Devereux87}, and for this pilot program we favored objects which had been detected in the past. 
The sample as a whole may therefore be biased towards galaxies which are bright in the IR. The LINERs in this sample are, however, ''IR-faint'' in terms of their FIR/optical luminosity ratios. Values of L$_{FIR}$/L$_{B}$ range from 0.1 - 5 for this sample, compared with median values of L$_{FIR}$/L$_{B}$ of 106 and 1.8 for the IR-bright and IR-faint objects in \citet{Sturm06}, respectively. IR-bright and IR-faint LINERs are cleanly separated in several MIR diagnostic diagrams, and may represent entirely different phenomena \citep{Sturm06}. Although many IR-bright LINERs show the compact hard X-ray sources characteristic of AGN activity \citep[e.g.][]{Dudik05}, the optical LINER emission in those objects may not be directly related to the accretion onto the central black hole \citep{Ho08}. 

There is evidence that each of the galaxies in the sample genuinely hosts an actively accreting, supermassive black hole. This evidence takes the form of an unresolved hard X-ray point source, optical/UV variability, high-ionisation MIR emission lines, and other phenomena. The specific evidence for each object is given in Table \ref{basic_data}. We estimate the Eddington ratios, $\rm L_{bol} / L_{Edd}$, of the AGNs using the stellar velocity dispersions of \citet{Ho09a}, the $\rm M_{BH} -\sigma$ relation of \citet{Tremaine02} and X-ray based bolometric luminosities. As expected, the Eddington ratios are low: $ 6 \times 10^{-7} < \rm L_{bol} / L_{Edd} < 4 \times 10^{-4}$. For comparison, \citet{Ho09} finds a median $\rm L_{bol} / L_{Edd} \sim 10^{-6}$ for the LINERs and $\rm L_{bol} / L_{Edd} \sim 10^{-4}$ for the Seyferts in the Palomar sample. 

In \S\ref{results} onwards we divide the LLAGN into three sets for the purpose of discussion:

I. At log $\rm L_{bol} / L_{Edd} < -4.6$, with PAH-dominated spectra and lacking a strong nuclear point source, are the {\bf ``host-dominated, low-Eddington ratio''} sources. 

II. Also at log $\rm L_{bol} / L_{Edd} < -4.6$ we highlight a group of LLAGN with strong radio emission \citep[log $\nu L_{\nu} \rm (5 \; GHz)$ / $L_{X} > -2.7$; cf][]{Terashima03} and well-sampled SEDs that allow strong constraints to be placed on the origin of their IR emission. These are the {\bf ``radio-loud, low-Eddington ratio''} objects. 

III. Galaxies with log $\rm L_{bol} / L_{Edd} > -4.6$ (the Eddington ratio of NGC~3718, which has the lowest value of the objects  -- aside from NGC~4486/M87 --  with a compact nucleus (\ref{img})), form the  {\bf ``high Eddington-ratio''} category. 

\begin{figure}[t]
\includegraphics[scale=0.48,  clip, trim=20 0 0 0]{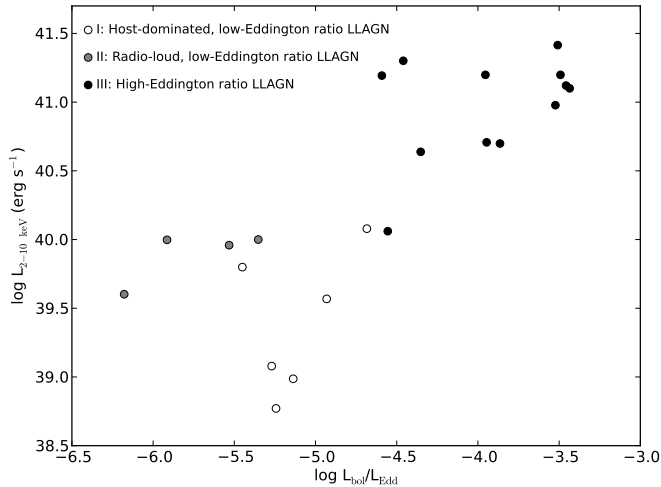}
\caption{ {\small Absorption-corrected hard X-ray luminosity vs Eddington ratio for the LLAGN sample. See \S\ref{sample} for definitions of the three LLAGN classes.}}
\label{L_vs_REdd}
\vspace*{7mm}
\end{figure}

We choose to discuss the LLAGN in terms of Eddington ratio as opposed to some other quantity (e.g. luminosity, morphology, spectral features) as there are indications that accretion rate is the fundamental parameter governing the properties of these objects \citep[e.g][]{Yuan07,Ho09,Elitzur09,Trump11,Yu11}. As shown in Figure \ref{L_vs_REdd}, Eddington ratio is loosely correlated with hard X-ray luminosity in this sample.

\begin{deluxetable*}{lcccccccc}
\tablecaption{Source Properties \label{basic_data}}
\tablewidth{0pt}
\tablehead{
\colhead{Galaxy} &  Morph. & \colhead {AGN Type\tablenotemark{a}}  & \colhead{Distance}  & \colhead{log $M_{\rm BH}$}& \colhead{log $\lambda$\tablenotemark{b}} & \colhead{$R_{X}$\tablenotemark{c}} & \colhead {AGN Evidence\tablenotemark{d,e}} & \colhead{Category\tablenotemark{f}}\\
 & & & \colhead{(Mpc)} &  \colhead{($M_{\odot}$)} & & & \colhead{(Ref) }}
\startdata

NGC~1052  & E4 & L1.9  & 19.4             & 8.3 & -4.0 & -1.5 & H$\alpha$ (1) & III \\
                       &       &           &                     &   && & Pol H$\alpha$ (2) \\
                      &         &  &                     &  && & X-ray (3, 4) \\
                      &         &  &                     & && & [NeV] (5)\\
                      &         &  &                     &  && & NLR (6)\\
NGC~1097  & SB(s)b & L1      & 14.5             & 8.1 & -4.4 & -4.0 &    H$\alpha$ (17) & III \\
	              &            & &                  &         && & X-ray (16) \\
NGC~3031   & SA(s)ab &  S1.5 & 3.6     & 7.8 &  -4.0 & -3.5 & H$\alpha$ (1) & III \\
(M81)	     &           &&                   &        &&& X-ray (3) \\
	              &           &&                   &        &&& UV (11) \\
	              &           &&                   &        &&& Radio (10) \\
NGC~3166   & SAB0/a(rs)  & L2     & 35.4            & 7.7 & -3.9 & ... & [NeV] (5) & III \\
NGC~3169   & SA(s)a & L2     & 19.7            & 8.0 & -3.5 & -4.3 & X-ray (7) & III \\
	              &           &&                   &      &  & & Radio (8, 10) \\
NGC~3718    & SB(s)a & L1.9  & 17.0            & 7.7 & -4.6 & -3.1 & X-ray (7) & III \\
		      &           &&                   &     &&& Radio (10) \\
		      &           &&                   &       & && H$\alpha$ (1)\\
NGC~3998     & SA(r)0 & L1.9  & 14.1            & 8.9 & -4.5 & -3.3 & H$\alpha$ (1) & III \\
		      &           &&                   &     &&& Radio (10) \\
	              &           &&                   &     &&& X-ray (18) \\
		      &           &&                   &     &&& UV (11)\\
NGC~4111    & SA(r)0 & L2      & 17.0            & 7.6 & -4.9 & ... & X-ray (4, 12) & I \\
NGC~4258    & SAB(s)bc & S1.9  & 7.2              & 7.6 & -3.5 & -5.3 & H$\alpha$ (1) & III \\
		      &           &&                  &&        && UV (11) \\
		      &           &&                   &&        && Radio (10) \\
NGC~4261    & E2-3 & L2     & 35.1 & 8.9 & -4.6 & -2.4 & X-ray (4, 7) & III \\
(3C 270)         &           &&                     & &&& Pol H$\alpha$ (2)\\
		      &           &&                     &&&& Radio (10)\\
NGC~4278    & E1-2 & L1.9  & 9.7                 & 8.6 & -5.5 & -2.6 &H$\alpha$ (1) & II \\
		      &           &&                     &&&& Radio (8, 10) \\ 
		      &           &&                     &&&& X-ray (3, 7, 9, 13, 18) \\
		      &           &&                     &&&& [NeV] (5) \\
NGC~4374    & E1 & L2     & 16.8       & 8.9 & -6.2 & -1.1 & Radio (10) & II \\
(M84)              &           &&                     &&&& X-ray (3, 4, 13) \\
NGC~4438    & SA0/a(s) & L1.9  & 16.8             & 7.5 & -5.3 & ... & H$\alpha$ (1) & I  \\
		      &           &&                     &    & && X-ray (12, 18) \\
		      &           &&                     &    & && [Ne V] (5) \\
NGC~4457    & (R)SAB0/a(s) & L2     & 17.4              & 7.2 & -5.1 & ... & X-ray (4, 7, 12) & I \\
NGC~4486    & cD0-1 & L2     & 16.0   & 9.0 & -5.9 & -1.5 & Radio (10) & II  \\
(M87, 3C 274) &          &&                  & &&& X-ray (3, 4) \\
NGC~4579 & SAB(rs)b & S1.9/L1.9 & 16.8 & 7.8 & -3.5 & -3.8 & X-ray (3, 4, 9) & III \\
(M58)           &           &&                      & &&& UV (11) \\	
		    &           &&                       & & && Radio (8, 10) \\
NGC~4594 & SA(s)a & L2   & 9.8      & 8.5 & -5.4 & -2.2 & X-ray (4, 13) & II \\
(M104)        &         &&                   & &&& UV (11) \\
NGC~4736 & (R)SA(r)ab & L2    & 4.3    & 7.1 & -5.2 & -3.2 & X-ray (4) & I \\
(M94)           &       &&               &&&& UV (11) \\
		      &       &&               &&&& [NeV] (5) \\
NGC~5005     & SAB(rs)bc & L1.9  & 13.7  & 7.9  & -4.7 & ... & H$\alpha$ (1) & I \\
		      &           &&                      & &&& X-ray (4,5) \\
NGC~5033    & SA(s)c & S1.5  & 18.7        & 7.6 & -3.4 & -4.3 & H$\alpha$ (1) & III \\
		     &           &&                      &&     && X-ray (9, 13) \\
		      &          &&                     &&    && [NeV] (15) \\
NGC~5363   & I0? & L2     & 22.4               & 8.4 & -5.5 & -1.9 & Radio (10) & I \\
	             &          &&                      &&&& X-ray (18) \\
NGC~7479   & SB(s)c & S1.9 & 32.4       & 7.7 & -3.5 & -3.9 & H$\alpha$ (1) & III \\
		     &          &&                      &&&& Radio (14) \\

\enddata
\tablenotetext{a}{Optical spectroscopic classification as given by \citet{Ho97}, except for NGC~1097. \citet{Phillips84} find that NGC~1097 has a ``classic'' LINER spectrum, and the detection of broad components to both the H$\alpha$ and H$\beta$ lines \citep{Storchi-Bergmann93} suggest that it should be classified as a type 1.}

\tablenotetext{b}{$\lambda \equiv \rm L_{bol} / L_{Edd}$}

\tablenotetext{c}{$R_{X} \equiv$ log $\nu L_{\nu} \rm (5 \; GHz)$ / $L_{X}$ }

\tablenotetext{d}{X-ray: Hard X-ray point source(s).  H$\alpha$: broad H$\alpha$ line. Pol H$\alpha$: broad H$\alpha$ detected in polarized light.  Radio: high brightness-temperature radio nucleus and/or radio jet. UV: UV variability. [Ne V]: [NeV] 24 $\mu$m line. NLR: ionization cone(s). This is not intended to be an exhaustive list for each object.}

\tablenotetext{e}{1: \citet{Ho97b}. 2: \citet{Barth99}. 3: \citet{Satyapal04}. 4: \citet{Gonzalez-Martin06}. 5: \citet{Dudik09}. 6: \citet{Pogge00}. 7: \citet{Satyapal05}. 8: \citet{Falcke00}. 9: \citet{Terashima03}. 10: \citet{Nagar05}.  11: \citet{Maoz05}; 12: \citet{Flohic06}.  13: \citet{Ho01}. 14: \citet{Laine08}. 15: \citet{Dudik07}. 16: \citet{Nemmen06}. 17: \citet{Storchi-Bergmann93}. 18: \citet{Gonzalez-Martin09a}.}

\tablenotetext{f}{I: Host-dominated, low-Eddington ratio galaxy (log $\rm L_{bol} / L_{Edd} <  -4.6$). II: Radio-loud, low-Eddington ratio galaxy (log $\rm L_{bol} / L_{Edd} < -4.6$, log $\nu L_{\nu} \rm (5 \; GHz)$ / $L_{X} > -2.7$). III: High-Eddington ratio galaxy (log $\rm L_{bol} / L_{Edd} > -4.6$). See \S\ref{sample} for details. }

\end{deluxetable*}

\section{Observations and Data Reduction}
\label{obsdr}

\subsection{Gemini Mid-Infrared Imaging}
\label{mir}

MIR imaging observations were performed using T-ReCS \citep{Telesco98} on the 8.1 m Gemini South telescope and Michelle \citep{Glasse97} on the Gemini North telescope. Both T-ReCS and Michelle use a Raytheon 320 $\times$ 240 pixel Si:As IBC array. This provides a plate scale of 0\arcsec.09 and 0\arcsec.10 per pixel respectively, corresponding to a field of view (FOV) of 28\arcsec\ $\times$ 21\arcsec\ for T-ReCS and 32\arcsec\ $\times$ 24\arcsec\ for Michelle. The standard chop-nod technique was used to remove time-variable sky background, telescope emission and 1/f detector noise. In all observations the chop throw was 15\arcsec, the chop angle was 30\deg\ E of N and the telescope was nodded approximately every 40 s.

The Si-2 (8.7 $\mu$m, $\Delta \lambda$ = 0.8 $\mu$m, 50\% cut-on/off) filter in T-ReCS provides the best combination of sensitivity and angular resolution, so this filter was used for the majority of the T-ReCS observations. NGC~1052 was also observed with T-ReCS' Qa (18.3 $\mu$m, $\Delta \lambda$ = 1.5 $\mu$m, 50\% cut-on/off) filter. Most of the Michelle observations were taken using the sensitive semi-broad N$^{\prime}$ filter (11.2 $\mu$m, $\Delta \lambda$ = 2.4 $\mu$m, 50\% cut-on/off). Some of the brighter galaxies were also observed with Michelle in one or more of the Si-2 (8.8 $\mu$m, $\Delta \lambda$ = 0.9 $\mu$m, 50\% cut-on/off), Si-4 (10.3 $\mu$m, $\Delta \lambda$ = 1.0 $\mu$m, 50\% cut-on/off), Si-6 (12.5 $\mu$m, $\Delta \lambda$ = 1.2 $\mu$m, 50\% cut-on/off) and Qa (18.1 $\mu$m, $\Delta \lambda$ = 1.9 $\mu$m, 50\% cut-on/off) filters. For galaxies with existing, published high-resolution MIR photometry, the Michelle filters were chosen to complement the available data. A summary of the observations, which were taken in queue mode\footnotemark, is given in Table \ref{mir-obs}. 

\footnotetext{Program IDs: GN-2007A-Q-93, GS-2007B-Q-203, GN-2008A-Q-43, GN-2011A-Q-55}
	
The data were reduced using the Gemini IRAF package and our own IDL routines. The difference for each chopped pair was calculated and the nod-sets then differenced and combined to create a single image. During this process all nods were examined for high background that could indicate the presence of clouds or high water vapor, but no data needed to be removed for this reason. In a few cases where AB pairs showed unusually high electronic noise, those pairs were rejected.
	
Observations of Cohen standards \citep{Cohen99} were obtained immediately before and/or after each galaxy observation using the same instrument configuration. 
A Moffat function with two free parameters, FWHM and $\beta$, best described the delivered PSF \citep[cf][]{Radomski08}. The galaxy images were flux calibrated using standard star brightnesses appropriate for the instrument, filter, airmass and standard star SED\footnote{ Calculated using the web form available at http://www.gemini.edu/?q=node/11246}. At this stage an artificial source of known flux (10 mJy) was introduced into the calibrated images, with the same FWHM and $\beta$ parameter as the associated standard star. This source was used to quantify changes to the signal-to-noise ratio (S/N), galaxy brightness and FWHM during the subsequent reduction steps, and to provide an estimate of the error on the photometric measurements.

Vertical and/or horizontal electronic patterns in the images were then removed, if necessary. This was achieved using the ``miclean'' task in the Gemini IRAF package, which sequentially models and removes electronic noise and background structures. 
Finally, in the cases where the galaxy was observed more than once, the data were combined. 

Two different photometric measurements were made for each galaxy (Table \ref{mir-phot}). First, the flux in a circular aperture of 2\arcsec\ diameter was measured. Second, the standard star was scaled to the centroid of the galaxy emission and then the flux measured in a 2\arcsec\ aperture around the scaled star. We refer to these methods as ``aperture'' and ``PSF-scaling'' photometry, respectively, in the remainder of the paper. The PSF-scaling photometry represents the maximum likely contribution from an unresolved nuclear source, while the difference between the aperture and PSF-scaling measurements gives the approximate contribution of extended emission surrounding the central source. Various authors have further refined the PSF-scaling technique by adjusting the scaling of the standard star to account for the underlying host galaxy emission that undoubtedly contributes at the nucleus \citep{Radomski03, Levenson09, RamosAlmeida09, RamosAlmeida11}. However, because of the low S/N in some of the images, and to provide a uniform set of measurements, we did not attempt this. The above groups typically find that in Seyfert galaxies, 70 -- 100\% of the simple PSF-scaling flux in the N band comes from the pointlike nucleus. We therefore expect the PSF-scaling photometry to provide a good estimate of the AGN brightness in LLAGN dominated by nuclear point sources. This will not be true for objects with very faint nuclei embedded in copious extended emission.

The uncertainty in the photometry was estimated in the following manner. First, the time-variability of the sky transparency was estimated from the variation in the signal from the standard stars, found to be $\sim$6\% at $\lambda \sim 10 \; \mu$m and $\sim$10\% at $\lambda \sim 18 \; \mu$m. Second, the difference between the initial and final fluxes of the artificial source was calculated for each galaxy. Third, for the PSF-scaling photometry, an estimate of the error induced by a variable PSF was obtained by cross-calibrating the standard stars observed on a single night. This error was found to be $\sim13$\%. The total uncertainly was calculated by adding in quadrature these individual contributions. The pixel-pixel variations in the residual sky background contribute a negligible amount to the photometric errors.

All galaxies but one, NGC~4374, were detected in the MIR imaging. We use the observations of NGC~4438, a faint object observed on the same night, to estimate a 3$\sigma$ upper limit on the 8.7~$\mu$m flux of NGC~4374. To complete the set of consistently obtained photometric measurements, we also re-analysed archival MIR imaging data for NGC~1097 \citep{Mason07} and NGC~4486 \citep{Perlman01}.

\begin{deluxetable*}{ccccccc}
\tablecaption{Summary of new MIR observations \label{mir-obs}}
\tablewidth{0pt}
\tablehead{
\colhead{Galaxy} & \colhead{Instrument} & \colhead{Date} & \colhead{Filter} & \colhead{Integration Time\tablenotemark{a}}  \\
                  & 			     &		        &		  &	\colhead{(s)}}
\startdata
NGC 1052	 & T-ReCS   &	20070817	    &  Si2  & 927      \\
		 & T-ReCS   &	20070820	    &  Qa  & 927      \\
NGC 3031& Michelle &    20110203      & Si2  & 311     \\	
                   & Michelle &  20110203       &  Si4  & 309        \\
                   & Michelle &  20110203      &  Si6   & 329     \\
                   & Michelle & 20110204      &    Qa   & 324     \\	 
NGC 3166 & T-ReCS   & 	20071202    & 	Si2 & 927   \\
NGC 3169 & T-ReCS   & 	20071203, 20071206  & 	Si2 & 1853 \\
NGC 3718	 & Michelle  & 	20081107	    &  N$^{\prime}$    & 855    \\
NGC 3998  & Michelle  &	20110123    & 	Si2 & 311    \\	      
                   & Michelle  &    20110123   & 	Si4 & 309   \\	                 
                   & Michelle  &	20081206    &	Si6 & 753    \\
                   & Michelle   &   20110123   & Qa   & 324     \\
NGC 4111	 & Michelle &	20081206    &	N$^{\prime}$   &	 743    \\
NGC 4258  &  Michelle & 20110203  &  Si2  &   311   \\
                   & Michelle  & 20110203  &  Si4  & 309    \\
                   & Michelle  & 20110203  &  Si6  & 329    \\
NGC 4261	 & T-ReCS &	20080103      &	Si2 & 927    \\
NGC 4278	 & Michelle &	20080616    &	N$^{\prime}$   &  525   \\
NGC 4374 & T-ReCS &    20080220  &    Si2     & 927    \\
NGC 4438	 & T-ReCS &	20080220, 20080704  &	Si2 & 1853  \\
NGC 4457 & T-ReCS & 	20080327	    &   Si2 & 927   \\
NGC 4579 &  Michelle & 20110205  & Si2  &  311   \\
                   & Michelle  & 20110123  & Qa  &  648   \\
NGC 4594 & T-ReCS & 	20080502       &	Si2 & 927   \\
NGC 4736	 & Michelle &	20070415    & 	N$^{\prime}$   &  612   \\
                    & Michelle &    20070416   &   Qa       &   512 \\
NGC 5005 & Michelle &	20081206    &	N$^{\prime}$   &	 900   \\
NGC 5033 & Michelle &	20080616    &	N$^{\prime}$   & 941   \\
NGC 5363	 & Michelle &	20081107	    &    N$^{\prime}$  & 900   \\
NGC 7479	 & Michelle &	20080701 &	Si2   & 933   \\
\enddata
\tablenotetext{a}{On-source integration time, excluding any rejected nods}

\end{deluxetable*}

\begin{deluxetable*}{ccccccccc}
\tablecaption{MIR photometric results \label{mir-phot}}
\tablewidth{0pt}
\tablehead{
\colhead{Galaxy} & \colhead{Filter}  & \colhead{F$_{2\arcsec}$\tablenotemark{a}} & \colhead{F$_{\rm PSF}$\tablenotemark{b}} & \colhead{FWHM$_{\rm gal}$\tablenotemark{c}} & \colhead{FWHM$_{\rm std}$\tablenotemark{d}}	 \\
      &         &  \colhead{(mJy)}  & \colhead{(mJy)}   & \colhead{\arcsec} & \colhead{\arcsec} }
\startdata
NGC 1052	& Si2 &    62  $\pm$ 4  		&    42 $\pm$ 7     		&	 0.42	     & 0.39  \\
		         & Qa  &    165 $\pm$ 17		&   ...\tablenotemark{e}  		&	 0.75     & 0.67\tablenotemark{f}  \\
NGC 1097\tablenotemark{g}  &  Si5 & 42 $\pm$ 3    &   39 $\pm$ 6         &       0.43     & 0.41   \\
                         & Qa  &    64 $\pm$ 4              &   63 $\pm$ 9                  &       0.54      & 0.52   \\		         
NGC 3031	& Si2 &    69  $\pm$ 5  		&   67 $\pm$ 10     		&	 0.37        &  0.36 \\
		         & Si4 &    145 $\pm$ 10 		&   140 $\pm$ 21  		 &	 0.39        & 0.35  \\
		         & Si6 &    129 $\pm$ 8 		&   115 $\pm$ 17  		 &	 0.43        &  0.39 \\
		         & Qa  &    257 $\pm$ 26          &  197  $\pm$ 32  	    	 &	 0.54        & 0.53  \\
NGC 3166	& Si2 &    7   $\pm$ 1 		&     3 $\pm$ 1  		 &	 \nodata     & 0.27  \\
NGC 3169	& Si2 &    13  $\pm$ 2 		&    9  $\pm$ 2   		&	 0.50	     & 0.46  \\
NGC 3718	& N$^{\prime}$   &    21  $\pm$ 2 		&    14 $\pm$ 3   		&	 0.35	     & 0.35  \\
NGC 3998	& Si2 &    27  $\pm$ 2 		&    26 $\pm$ 4   		&	 0.42	     & 0.33  \\	
			& Si4 &    57  $\pm$ 5 		&    42 $\pm$ 8   		&	 0.39	     & 0.34  \\	
			& Si6 &    70  $\pm$ 5 		&    52 $\pm$ 8   		&	 0.54	     & 0.46  \\
			& Qa  &    108 $\pm$ 11 		&   96 $\pm$ 16   		&	 0.59\tablenotemark{f}  &  0.53 \\
NGC 4111	& N$^{\prime}$   &    10  $\pm$ 2 		&    4  $\pm$ 2   		&	 \nodata     & 0.35  \\
NGC 4258	& Si2 &    64  $\pm$ 4 		&    64 $\pm$ 10 		 &	 0.32        & 0.31  \\
			& Si4 &    87 $\pm$ 6 		&    70  $\pm$ 11  		&	 0.33        & 0.32  \\
			& Si6 &    114 $\pm$ 8 		&   108 $\pm$ 16 		 &	 0.37        & 0.37  \\	
NGC 4261	& Si2 &    5   $\pm$ 1 		&     3 $\pm$ 1  		 &	 \nodata     & 0.48  \\
NGC 4278	& N$^{\prime}$   &    5   $\pm$ 1 		&     2 $\pm$ 1   		&	 \nodata     & 0.35  \\
NGC 4374	& Si2 &  $<$ 3          			 &     \nodata	     		&	 \nodata     & 0.32  \\
NGC 4438	& Si2 &    11  $\pm$ 1 		&    7  $\pm$ 2   		&	 0.33        & 0.32  \\
NGC 4457	& Si2 &    12  $\pm$ 1 		&    6  $\pm$ 1   		&	 0.47	     & 0.27  \\
NGC 4486\tablenotemark{g}  & N    & 17 $\pm$ 1  & 16 $\pm$ 3              &       0.46      & 0.46  \\
NGC 4579	& Si2 &    32  $\pm$ 2 		&    28 $\pm$ 4  		&	 0.29        & 0.24  \\
			& Qa  &    112 $\pm$ 12 		&   97 $\pm$ 16  		&	 0.58        & 0.52  \\
NGC 4594	& Si2 &    8   $\pm$ 2 		&     2 $\pm$ 1   		&	 \nodata     & 0.38  \\
NGC 4736	& N$^{\prime}$  &    33  $\pm$ 4 		&    11 $\pm$ 4   		&	 0.56	     & 0.48  \\
			& Qa  &    101 $\pm$ 11 		&   43  $\pm$ 8   		&	 \nodata     & 0.65  \\
NGC 5005	& N$^{\prime}$   &    26  $\pm$ 2 		&    4  $\pm$ 1   		&	 0.65        & 0.35  \\
NGC 5033	& N$^{\prime}$   &    22  $\pm$ 2 		&    15 $\pm$ 3   		&	 0.35	     & 0.35  \\
NGC 5363	& N$^{\prime}$   &    11  $\pm$ 3 		&    3  $\pm$ 2  		 &	 0.71        & 0.35  \\
NGC 7479	& Si2 &    158 $\pm$ 10 		&   157 $\pm$ 23  		&	 0.40	     & 0.40  \\
\enddata
\tablenotetext{a}{Photometry in a 2\arcsec\ diameter aperture}
\tablenotetext{b}{PSF-scaling photometry; see text}
\tablenotetext{c}{FWHM of nucleus}
\tablenotetext{d}{FWHM of standard star}
\tablenotetext{e}{For NGC~1052 only, the standard star image in the Qa filter contains a prominent diffraction ring. This means that the Moffat function used to model the PSF for the PSF-scaling photometry does not provide a good representation of the flux profile in this case.}
\tablenotetext{f}{Gaussian FWHM; Moffat function did not converge for this source}
\tablenotetext{g}{Re-analysis of data originally published by \citet{Perlman01}, \citet{Mason07}}
\end{deluxetable*}

\subsection{Gemini Near-Infrared Imaging}
\label{nir}

NIR imaging observations (GN-2011A-Q-55) were performed in queue mode on 20110204 using the Near InfraRed Imager and Spectrometer \citep[NIRI; ][]{Hodapp03} on Gemini North. The NIR observations were obtained for a subset of the brighter galaxies in the sample and were intended to fill in parts of the SEDs lacking in published data.

NIRI uses a 1024 x 1024 ALADDIN InSb array. With the f/32 camera, used to avoid saturation on the bright sky background, this provides a plate scale of 0.022\arcsec\ per pixel and FOV of 22.5\arcsec\ $\times$22.5\arcsec. Observations were taken through the L$^{\prime}$ (3.9 $\mu$m, $\Delta \lambda =$ 0.4 $\mu$m, 50\% cut-on/off ) and M$^{\prime}$ (4.8 $\mu$m, $\Delta \lambda =$ 0.1 $\mu$m, 50\% cut-on/off ) filters. Total integration times of 180 sec (L$^{\prime}$) and 225 sec (M$^{\prime}$) were used. The galaxy images were acquired in an ABB$^{\prime}$A$^{\prime}$ pattern, nodding approximately 50\arcsec\ to blank sky. 

The data were reduced using the Gemini IRAF package and standard NIR imaging procedures. The images were corrected for nonlinearity, sky-subtracted, and divided by a flat field constructed from sky frames. A bad pixel mask was also applied. The dithered observations were then shifted to a common position and averaged, and flux calibration was achieved using standard stars observed immediately before or after each galaxy \citep{Leggett03}. 

As with the MIR observations (\S\ref{mir}), an artificial source was introduced at this stage and both aperture and PSF-scaling photometry performed (Table \ref{nir-phot}).  
For NGC~3031 and NGC~4258 at M', the FWHM of the standard star is larger than that of the AGN, causing the PSF-scaling flux to be higher than that given by the aperture photometry. Therefore no PSF-scaling flux is quoted for these objects. The uncertainty in the photometry was calculated in the same way as for the MIR photometry (\S\ref{mir}).

NGC~4486 was not detected in these short M$^{\prime}$ band exposures. We use the observations of NGC~4258 to estimate a 3$\sigma$ upper limit on the M$^{\prime}$ flux density of NGC~4486.	

\begin{deluxetable*}{ccccccc}
\tablecaption{NIR photometric results \label{nir-phot}}
\tablewidth{0pt}
\tablehead{
\colhead{Galaxy} & \colhead{Filter} & \colhead{F$_{2\arcsec}$\tablenotemark{a}} & \colhead{F$_{\rm PSF}$\tablenotemark{b}} & \colhead{\rm FWHM$_{\rm gal}$\tablenotemark{c}} & \colhead{FWHM$_{\rm std}$\tablenotemark{d}}	 \\
      &        &  \colhead{(mJy)}  & \colhead{(mJy)}   & \colhead{(\arcsec)} & \colhead{(\arcsec)} }
\startdata
NGC3031  &   L'       &  74 $\pm$ 2     &  56 $\pm$ 5           &   0.42      & 0.40  \\
                     & M'      &  56 $\pm$ 4      &  \nodata                &  0.38  & 0.40  \\     
NGC3998  & L'          &  24 $\pm$ 2    &   11 $\pm$ 2               &  0.48  & 0.27  \\
                    & M'       &  18 $\pm$ 3              &  16 $\pm$ 3               & 0.39   & 0.38 \\
NGC4258   & L'         &    29 $\pm$ 2             &   25 $\pm$ 3           &   0.47   & 0.42  \\
                     & M'       &  23 $\pm$ 2                 &  \nodata           & 0.28   & 0.54  \\
NGC4486   & L'         &   8 $\pm$ 1              &  8 $\pm$ 1                 &  0.43   &0.35 \\
                     &  M'       &  $<$ 8                      &  \nodata                                 &  \nodata       &    0.50  \\
NGC 4579\tablenotemark{e} & L'    &    17 $\pm$ 2                      &   13 $\pm$ 2  &   0.65  &      0.35      \\
\enddata
\tablenotetext{a}{Photometry in a 2\arcsec\ diameter aperture}
\tablenotetext{b}{PSF-scaling photometry; see text}
\tablenotetext{c}{FWHM of nucleus}
\tablenotetext{d}{FWHM of standard star}
\tablenotetext{e}{The M$^{\prime}$ images of NGC~4579 showed a strong background gradient that made reliable photometry impossible}
\end{deluxetable*}

\subsection{Spitzer IRS spectra}

Archival low-resolution spectra taken with the {\em Spitzer Space Telescope}'s InfraRed Spectrograph \citep[IRS; ][]{Houck04} are available for 18 of the 22 LLAGN and were downloaded from the {\em Spitzer} science archive\footnotemark. Starting from the pipeline-processed ``bcd'' data, the files were background-subtracted either by subtracting spectra in different nod positions, or by subtracting sky observed in one order when the object was being observed in the other order, depending on the available observations. One-dimensional spectra were extracted using the point source extraction algorithm provided by the SPICE software package, which provides a wavelength-dependent aperture of 7.2\arcsec\ at 6 $\mu$m. A few objects were observed in spectral mapping mode; for these only the spectrum at the nuclear position was extracted. Inspection of the spatial profiles of the spectra showed that many of the galaxies are extended at Spitzer's angular resolution, to varying degrees. The slit loss correction applied by the SPICE point source flux calibration is appropriate for an unresolved source and cannot account for wavelength-dependent extended structure. This may affect the accuracy of the flux calibration and overall continuum slope in the spectra of the extended objects in this sample. \citet{Gallimore10} find errors of $\sim$10\% from this effect in their analysis of the 12 $\mu$m AGN sample (containing HII region galaxies, LINERs and Seyferts). Nonetheless, we expect the IRS spectra presented here to provide a useful overview of the general IR spectral characteristics of the LLAGN.

\footnotetext{A reduced IRS spectrum of NGC~3998, including data from the SL and LH modules, was kindly provided by E. Sturm \citep{Sturm05}.}

\section{Results}
\label{results}

\subsection{IR Morphology and the MIR/X-ray Relation}
\label{img}

\begin{figure*}
\hspace*{8mm}
\includegraphics[scale=0.9, clip, trim=40 370 0 80]{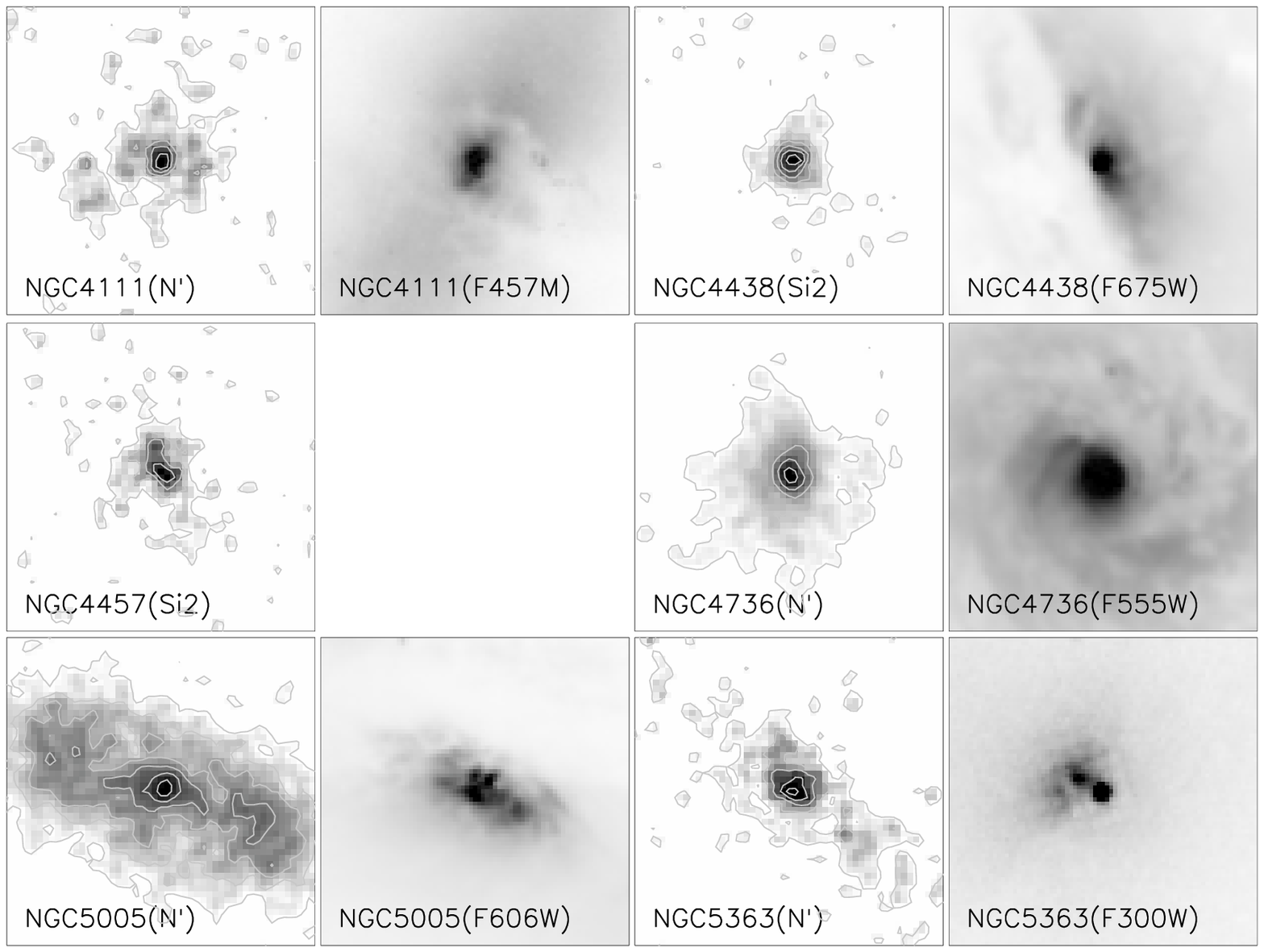}
\caption{ {\small Gemini MIR and HST optical/UV images of the category I, host-dominated LLAGN with log $\rm L_{bol} / L_{Edd} < -4.6$ (no HST image is available for NGC~4457). North is up and East left on the images, and a 5\arcsec\ $\times$ 5\arcsec\ region is shown.  Contours start at 1$\sigma$ and increase in steps of 1$\sigma$. All the MIR images were smoothed using a 2-pixel Gaussian.
 }}
\label{img1}
\end{figure*}

\begin{figure*}
\hspace*{8mm}
\includegraphics[scale=0.9, clip, trim=40 520 50 50]{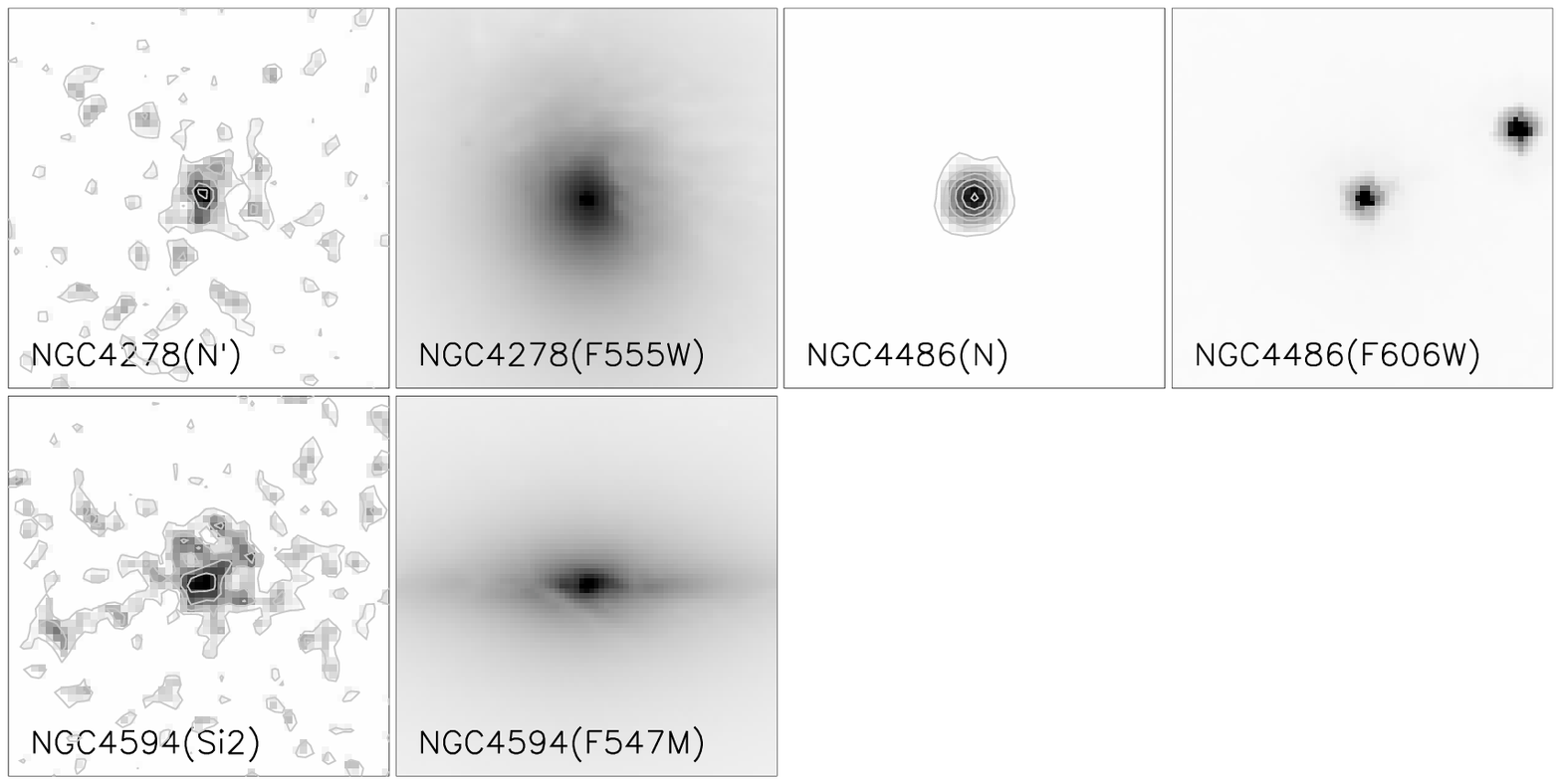}
\caption{ {\small Gemini MIR and HST optical images of the category II, radio-loud LLAGN with log $\rm L_{bol} / L_{Edd} < -4.6$.  NGC~4374, which was not detected in the MIR, also belongs in this category. Contours for NGC~4486 start at 10$\sigma$ and increase in steps of 10$\sigma$, others as in Figure \ref{img1}.}}
\label{img2}
\end{figure*}

\begin{figure*}
\hspace*{8mm}
\includegraphics[scale=0.85, clip, trim=30 250 0 0]{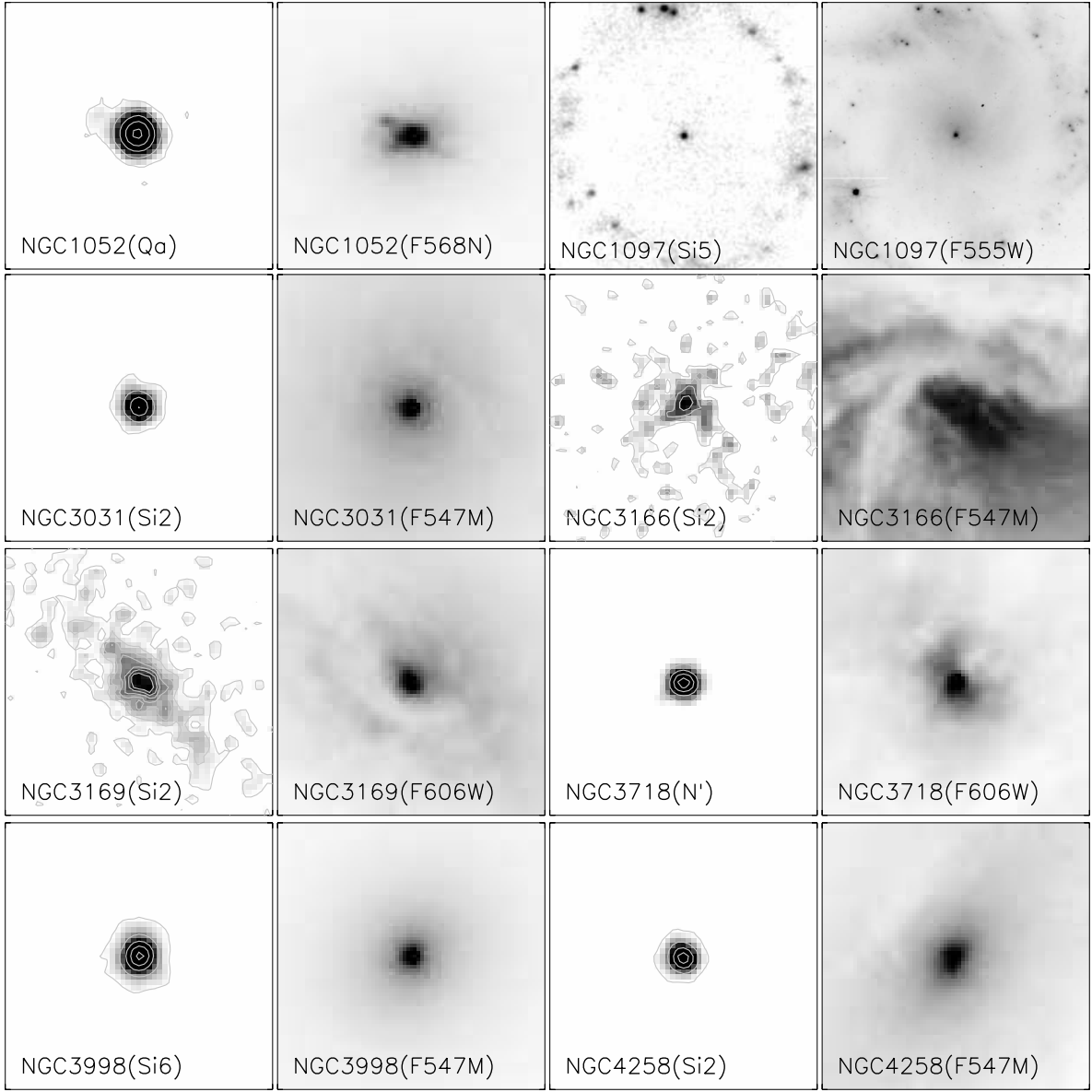}
\caption{ {\small Gemini MIR and HST optical images of the category III, high-Eddington ratio LLAGN (log $\rm L_{bol} / L_{Edd} > -4.6$). A 5\arcsec\ $\times$ 5\arcsec\ region is shown for every object except NGC~1097 (24\arcsec\ $\times$ 24\arcsec). Some of these galaxies were also observed in filters other than those shown here, but their morphologies are  very similar in all filters. Contours for NGC~1052 and NGC~3718 start at 5 $\sigma$ and increase in steps of 5 $\sigma$, contours for NGC~3998 start at 7 $\sigma$ and increase in steps of 7 $\sigma$, those for NGC~3031 and NGC~4258 start at 10 $\sigma$ and increase in steps of 10$\sigma$, others as in Fig. \ref{img1} }}
\label{img3}
\end{figure*}

\begin{figure*}
\hspace*{8mm}
\includegraphics[scale=0.83, clip, trim=30 500 0 0]{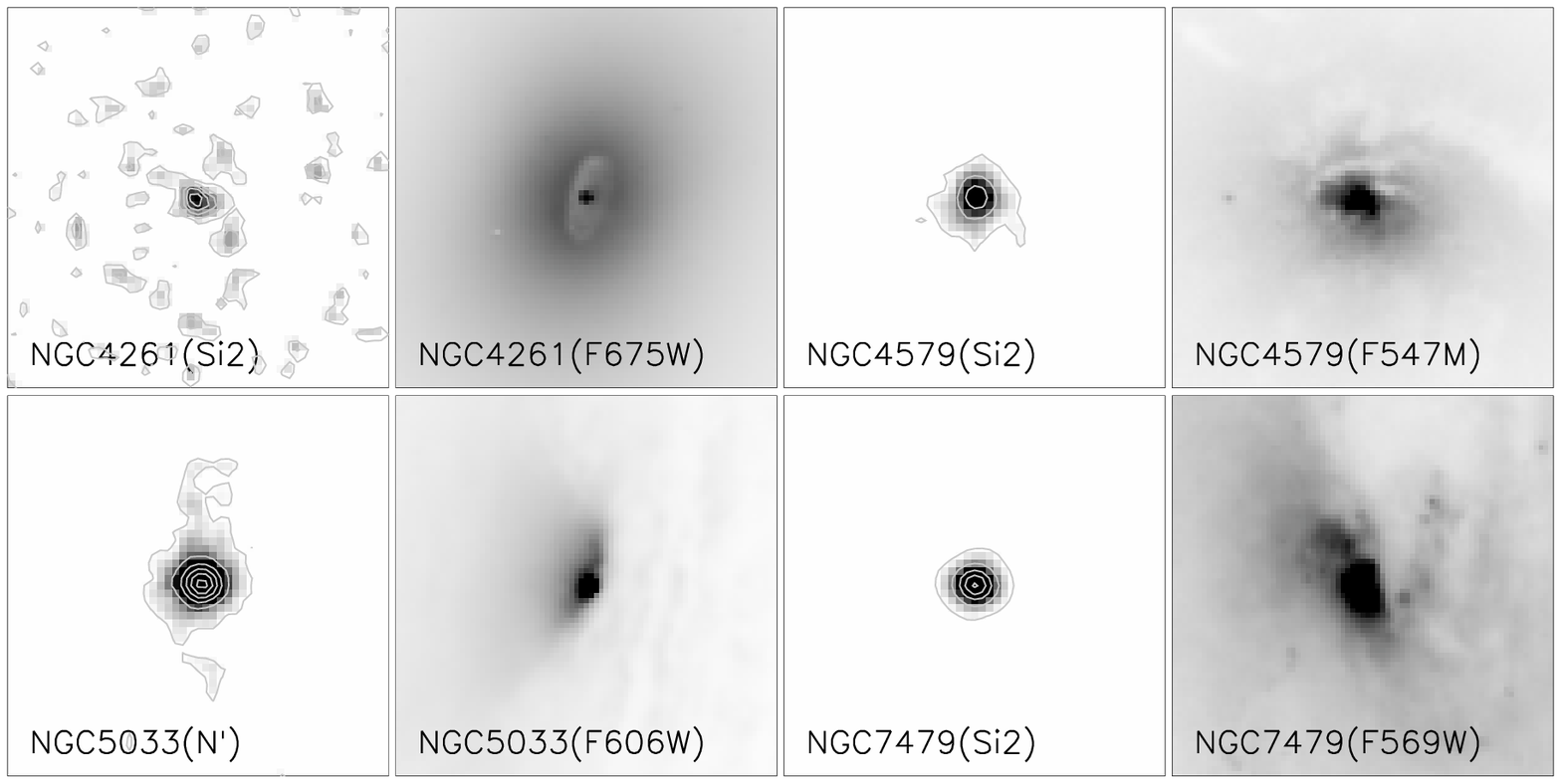}
\caption{ {\small High-Eddington ratio LLAGN, continued. Contours for NGC5033 start at 5$\sigma$ and increase in steps of 5$\sigma$, contours for NGC~4579 start at 7$\sigma$ and increase in steps of 7$\sigma$, and those for NGC7479 start at 40$\sigma$ and increase in steps of 40$\sigma$. NGC~4261 is as in Fig. \ref{img1}}}
\label{img4}
\end{figure*}

The MIR images are presented in Figures \ref{img1} -- \ref{img4}. For comparison, archival HST images in various optical/UV filters are also shown, where available.  The MIR filters used in this study have moderate bandwidths and encompass a number of possible emission and absorption features. Perhaps most significantly, the Si-2 and N$^{\prime}$ filters may contain PAH emission, while the Si-4 and Qa filters can be affected by silicate emission/absorption, if present. The [Ne II] fine structure line lies within the bandpass of the Si-6 filter. However, clear differences in morphology are not seen in objects observed in more than one filter and only one filter per object is shown in Figures \ref{img1} - \ref{img4}.

The images show a wide range of MIR morphologies. Some of the nuclei are Seyfert-like, in the sense that they are dominated by compact sources with FWHM comparable to that of the corresponding standard star. All of the galaxies observed at 3 -- 5~$\mu$m fall into this category and they all show strong, compact nuclei at those wavelengths as well. Others have a weaker nuclear MIR source embedded in a substantial amount of diffuse, extended emission. In some cases, notably NGC~5005 (Figure \ref{img1}), the MIR emission extends over several arcseconds ($>$100 pc). 

In the type 1.5 Seyfert NGC~5033 (Figure \ref{img4}), faint emission regions are detected to the north and south of the nucleus, approximately coincident with the $\rm [O III]$ emission observed in this object at lower angular resolution \citep{Mediavilla05}. This is likely one of the relatively few Seyferts in which the narrow-line region (NLR) has been resolved in the MIR. The 18~$\mu$m image of NGC~1052 (Figure \ref{img3}) shows faint emission along the radio jet axis, also detected in optical H$\alpha$+[N II] images \citep{Pogge00}. However, this image does show some low-level trefoil structure and the slight extension observed in the 8.7~$\mu$m image is offset by about 10\deg\ from the 18~$\mu$m extended emission, so detection of the NLR/jet in NGC~1052 should be regarded as tentative.

\subsubsection{Trends in morphology with Eddington ratio/luminosity}
\label{morph}

Figures \ref{img1} - \ref{img4} demonstrate differences in morphology between objects with low and high Eddington ratios. Strong nuclear point sources are common,  although not universal, in the high-$\rm L_{bol}/L_{Edd}$ objects. Apart from NGC~4486 (which is pointlike) and NGC~4374 (which we did not detect), all of the objects in the low-$\rm L_{bol}/L_{Edd}$ categories have nuclei that are much weaker compared to the surrounding diffuse emission. As discussed in \S\ref{intro}, the presence or absence of a strong nuclear IR source could be related to the presence or absence of a dusty torus. However, as illustrated by Figure \ref{L_vs_REdd}, low-$\rm L_{bol}/L_{Edd}$ objects tend also to have relatively low X-ray luminosities, which could mean that a nuclear point source is simply too faint to detect and/or is overwhelmed by host galaxy emission in these objects. We now quantitatively investigate this point.

To roughly quantify the extent to which each galaxy is dominated by a point source, we take the ratio of the PSF-scaling photometry and the 2\arcsec\ aperture photometry (\S\ref{obsdr}; Table \ref{mir-phot}). The resulting values range from 0.15 in NGC~5005  to 1.0 in NGC~4258. We refer to galaxies with $F_{PSF}/F_{2\arcsec\ } > 0.65$ as ``point-source-dominated'', or ``compact'', and the remainder of the objects as ``diffuse''.  This division is arbitrary, but it corresponds quite well to our subjective, visual classification of objects having a prominent nuclear source. 

The well-known MIR/X-ray relation can be used to predict whether the torus should be detectable in this shallow imaging survey. In Seyfert galaxies, high-energy photons are reprocessed into IR radiation by dust in the immediate surroundings of the active nucleus.  As a result of this interaction, a fairly tight correlation has been shown to exist between the MIR and hard X-ray emission in these objects \citep[with the X-rays essentially acting as a proxy for optical/UV emission;][]{Krabbe01,Lutz04,Horst06,Horst08,Ramos07,Gandhi09,Hardcastle09,Levenson09,Honig10a}. As the SEDs of LLAGN may be deficient in UV and optical radiation \citep[][but see also Eracleous et al. 2010]{Ho99}, the Seyfert-based MIR/X-ray relation may overestimate the reprocessed MIR luminosity expected from lower-luminosity objects. However, we use the relation as a starting point to assess the detectability of the torus. We fit the combined sample of the ``well-resolved'' objects in \citet{Gandhi09} and the ``PSF-fitting'' photometry of \citet{Levenson09}, measurements which are intended to most accurately represent the nuclear emission of Seyferts and quasars, finding $\rm log \; L_{X} = 0.88 \; log \; L_{MIR} + 4.75$. The fit is then extrapolated to the luminosities of the LLAGN.

\begin{figure}[t]
\includegraphics[scale=0.5, clip, trim=35 0 0 0]{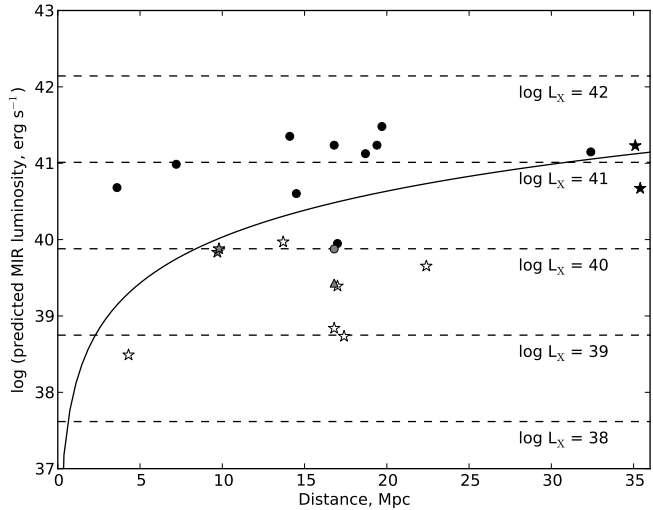}
\caption{ {\small Detectability of the torus in the LLAGN sample, based on the MIR luminosity predicted by the MIR/X-ray relation (Figure \ref{mirx_all}) and assuming a limiting sensitivity of 3 mJy. Objects above the solid line are expected to have detectable torus emission. Black symbols denote high-$\rm L_{bol} / L_{Edd} $ objects  (category III), while radio-loud and host-dominated low-$\rm L_{bol} / L_{Edd}$ nuclei (I, II) are shown by gray and white symbols, respectively. Galaxies found to have compact MIR morphologies are given circular symbols, while diffuse galaxies are shown with star symbols. NGC~4374, which was not detected and whose morphology could therefore not be determined, is shown as a triangle.}}
\label{detect}
\end{figure}

Figure \ref{detect} shows that the compact, relatively high-luminosity, high-$\rm L_{bol} / L_{Edd}$ sources are predominantly those in which the torus is expected to be detectable in this imaging survey\footnotemark. That we observe strong, unresolved MIR emission in the objects in which we would expect to detect the torus is consistent with most or all of that emission arising in the torus. Other AGN-related emission mechanisms may well be important, though: we also find that NGC~4486 and NGC~3718, in which the torus should not be detectable, host strong, pointlike nuclei. We may expect that the IR emission from those point-source-dominated objects is associated with the AGN rather than the host galaxy, but contains a significant contribution from sources other than the torus. We return to this point in \S\ref{SED_char}.

\footnotetext{This simple calculation does not take into account the difficulty of isolating a faint torus surrounded by relatively bright host galaxy emission, but for the purposes of interpreting Figure \ref{detect}, that is only important for galaxies just above the threshold.}

Conversely, the MIR/X-ray relation predicts that torus emission in most of the diffuse objects should be too faint to be detected. While the objects with diffuse morphologies mainly come from the low-Eddington ratio categories, the lack of a dominant pointlike nucleus can be explained simply by the lower luminosity of the AGN relative to the host galaxy. The image morphology alone, then, is not sufficient to demonstrate trends of torus properties with Eddington ratio.

\subsubsection{The position of the LLAGN on the MIR/X-ray relation}
\label{mirxsect}

The location of the LLAGN relative to the MIR/X-ray relation introduced in the previous section will depend on the X-ray data chosen to represent each galaxy. Several X-ray luminosities have been published for some of the objects, sometimes with rather different values. This was carefully considered during the course of this work, and the selection of the intrinsic X-ray luminosities is discussed in \S\ref{sect:seds}.

\begin{figure}
\hspace*{-8mm}
\includegraphics[scale=0.5, clip, trim=0 0 0 0]{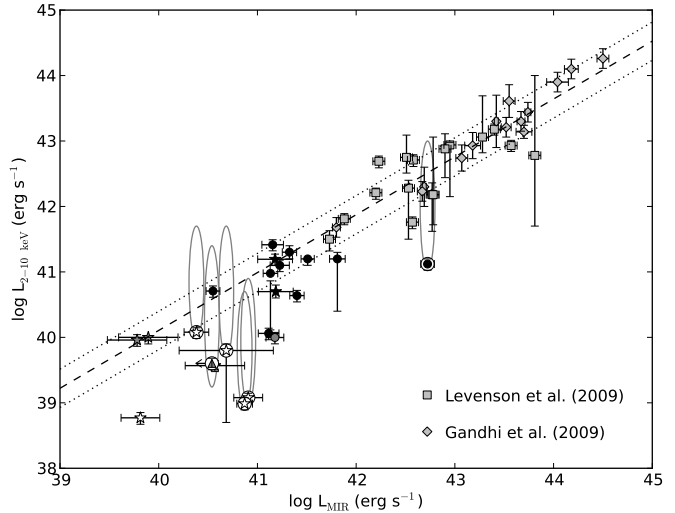}
\caption{Hard X-ray vs MIR luminosities for the LLAGN and for the higher-luminosity AGN studied by \citet{Gandhi09} and \citet{Levenson09}.  The dashed line shows the fit for the combined Gandhi and Levenson samples, extrapolated to lower luminosities, and the dotted lines mark the $+/-1\sigma$ scatter in the Gandhi/Levenson data. Errors on $\rm L_{X}$ are as stated in the references in Table \ref{measurements}, otherwise assumed to be $\pm$20\%. Ellipses show the range of available intrinsic X-ray luminosity estimates for possible Compton-thick objects, which are identified by encircled markers. Other symbols and colors for the LLAGN as in Figure \ref{detect}.}
\label{mirx_all}
\end{figure}

Figure \ref{mirx_all} shows that the diffuse, host-dominated low-$\rm L_{bol} / L_{Edd}$ objects, which predominate at the low-luminosity end of the sample, tend to have higher MIR/lower X-ray luminosities than predicted by the MIR/X-ray relation. Four of the six host-dominated, low-$\rm L_{bol} / L_{Edd}$ objects (NGC~4438, NGC~4457, NGC~5005 and NGC~5363) are suspected by \citet{Gonzalez-Martin09a} to be Compton-thick (Table \ref{measurements}). This could contribute to an apparent MIR excess/X-ray deficit if the intrinsic X-ray luminosity is not determined accurately. For the suspected Compton-thick nuclei, Table \ref{lx_ct} gives intrinsic luminosity estimates obtained from (a) spectral analysis of high-resolution, $\sim$few keV X-ray data, (b) the ratio of $\rm F_{X}(2-10 \; keV)/F ([O III])$  \citep{Gonzalez-Martin09a,Panessa06}, and (c) the $\rm L_{bol, AGN} - L _{[O IV]}$ relation \citep{Goulding10,Pereira-Santaella10}. The resulting range of estimated intrinsic luminosities for the Compton-thick nuclei is indicated in Figure \ref{mirx_all}. In all cases, when the possible large obscuring column is taken into account, the AGN could be luminous enough that the nuclei are consistent with the MIR/X-ray relation, or show a MIR deficit.

Strong PAH emission (\S\ref{irs}), which lies in the filters used to image these galaxies, may enhance the host galaxy emission relative to the rest of the sample. It may also indicate a contribution from dust heated by circumnuclear star formation. The PAH emission in the relatively large-aperture IRS spectra accounts for about 20 -- 40\% of the flux in the Michelle/T-ReCS filter bandpasses in the host-dominated LLAGN. This is probably an upper limit to the PAH contribution in the ground-based images, as the PAH emission is expected to be more extended than that of the AGN \citep[e.g.][]{Diaz-Santos10,Honig10a,Asmus11}. The PAH emission alone, then, cannot account for the entire MIR excess. As noted in \S\ref{mir}, though, even the combination of high-resolution imaging and PSF-scaling photometry is unlikely to effectively isolate the weak AGN emission in these diffuse objects. Some of the emission must arise in the stellar population of the host galaxy; in the photospheres of cool stars and in dusty circumstellar shells, for example. The contribution of each of these processes is difficult to assess reliably with the current data. It is possible, however, that once taken into account, the AGN MIR emission may be weaker than expected from the MIR/X-ray relation, especially if the nuclei are heavily obscured.

Finally, we note that emission from X-ray binaries may be important  in very low-luminosity objects ($\rm L_{X} \lesssim 10^{39} \rm \; erg \; s^{-1}$). While this would not explain the excess MIR emission, it does further emphasise that the location of very low-luminosity AGN on the MIR/X-ray plot may be influenced by a complex mix of factors.

Of the four radio-loud, low-$\rm L_{bol} / L_{Edd}$ galaxies, NGC~4278 and NGC~4594 have MIR luminosities consistent with the MIR/X-ray relation, while NGC~4486 has a large MIR excess and NGC~4374 (a suspected Compton-thick object) has only a MIR upper limit. 
The images of NGC~4278 and NGC~4594 suggest a significant host galaxy contribution to the PSF-scaling photometry. Furthermore, in \S\ref{SED_char} we argue that synchrotron radiation dominates the MIR emission in these nuclei. Therefore, despite the fact that they are ostensibly consistent with the MIR/X-ray relation for Seyfert galaxies, it is unlikely that NGC~4278 and NGC~4594 host a Seyfert-like torus. We also find (in \S\ref{SED_char}) that dust emission contributes little to the MIR luminosity of NGC~4486.  However, the MIR/X-ray relation predicts that reprocessed IR emission from a torus in NGC~4486 would account for $<10\%$ of the observed emission. We cannot estimate the amount of nonthermal MIR emission accurately enough to rule out torus emission at this level, so taken in isolation the MIR photometry remains consistent with the presence of a torus in this LINER.

A MIR deficit in the MIR/X-ray plot would be consistent with suggestions that a Seyfert-like torus does not exist in the LLAGN. However, the remaining objects, those in the high-luminosity, high-$\rm L_{bol} / L_{Edd}$ category, lie close to the MIR/X-ray relation or exhibit excess MIR emission. In most cases the MIR excess is only significant at the 1-2$\sigma$ level, but it is noteworthy that only two  of the 13 high-$\rm L_{bol} / L_{Edd}$ objects lie to the MIR deficit side of the fit. Similar results are reported by \citet{Asmus11}; while finding that the MIR/X-ray relation is formally unchanged down to L$\sim10^{41} \rm erg \; s^{-1}$, their LLAGN are offset from the relation by about 0.3 dex to the MIR excess side.

 \citet{Horst08} find that, even in a sample of Seyfert galaxies with pointlike nuclei in high-resolution MIR imaging, objects with FWHM $ \rm >560 \; \times \rm r_{sub}$ (the dust sublimation radius) are systematically offset from the MIR/X-ray relation for ``well-resolved'' AGN. They attribute this to contamination from nuclear MIR sources other than the torus. All of the LLAGN in this paper have FWHM $ \rm >560 \; r_{sub}$, so even the high-luminosity, high-$\rm L_{bol} / L_{Edd}$ with compact nuclei may be subject to contamination. The PSF-scaling photometry used in \S\ref{obsdr} (compared to the aperture photometry of Horst et al.) should mitigate this to some extent, but a contribution from unresolved, non-torus emission sources -- nuclear star formation, synchrotron radiation, truncated accretion disk, etc. -- is possible. We return to this issue in \S\ref{discuss}. 

Finally, NGC~7479 and NGC~3718, two objects with compact nuclei and very large MIR excesses, deserve special mention. The very strong silicate absorption feature in NGC~7479 (\S\ref{irs}), unique among this LLAGN sample, signifies that this nucleus is deeply embedded in a continuous medium of high optical depth \citep{Sirocky08}. The nucleus may be Compton thick and the intrinsic AGN luminosity underestimated (Table \ref{lx_ct}; \S\ref{sect:seds}). \citet{MunozMarin07} find the nucleus of NGC~7479 to be slightly resolved in HST UV imaging, so a nuclear star cluster could perhaps also contribute to the dust heating and therefore to the MIR luminosity of this object. In the case of NGC~3718, we later present evidence that some of the nuclear IR emission is produced by synchrotron radiation from the jet (\S\ref{RQSED}).

\subsection{5 -- 35 $\mu$m Spectra}
\label{irs}

\begin{figure*}[htpb]
\includegraphics[scale=0.9]{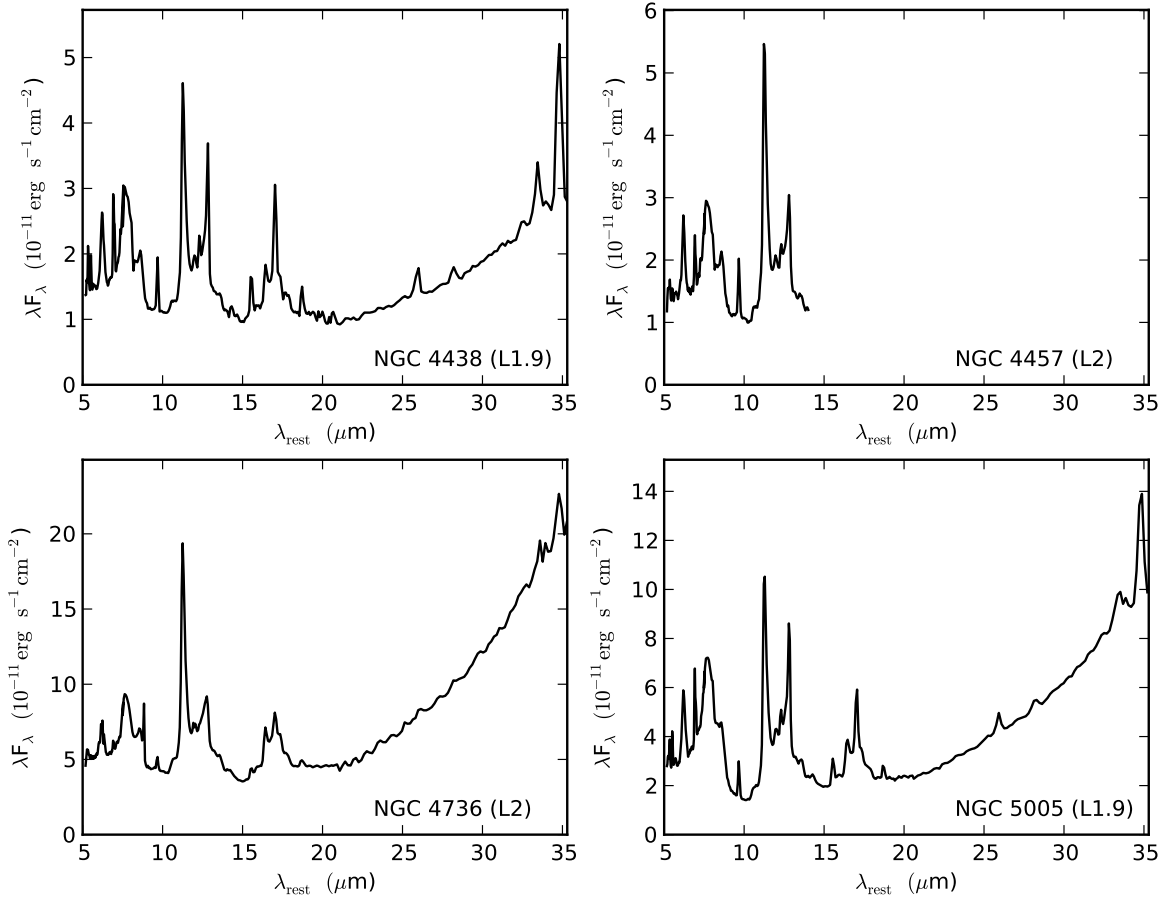}
\caption{ {\small MIR spectra of category I, host-dominated LLAGN with log $\rm L_{bol} / L_{Edd} < -4.6$}}
\label{irs1a}
\end{figure*}

\begin{figure}[htpb]
\includegraphics[scale=0.9, clip, trim=30 200 290 0]{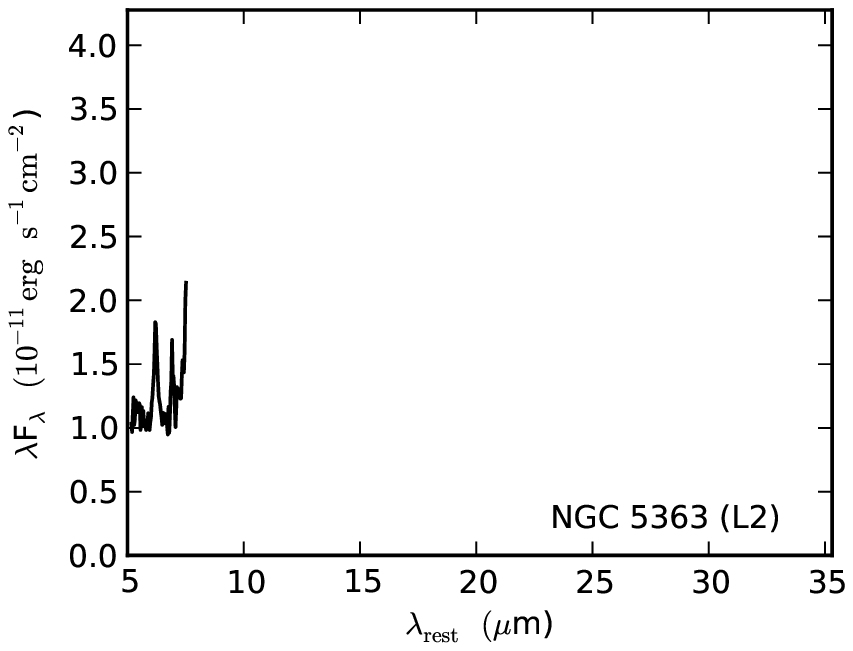}
\caption{ {\small MIR spectra of category I, host-dominated LLAGN with log $\rm L_{bol} / L_{Edd} < -4.6$, continued.}}
\label{irs1b}
\end{figure}

\begin{figure*}[htpb]
\includegraphics[scale=0.9]{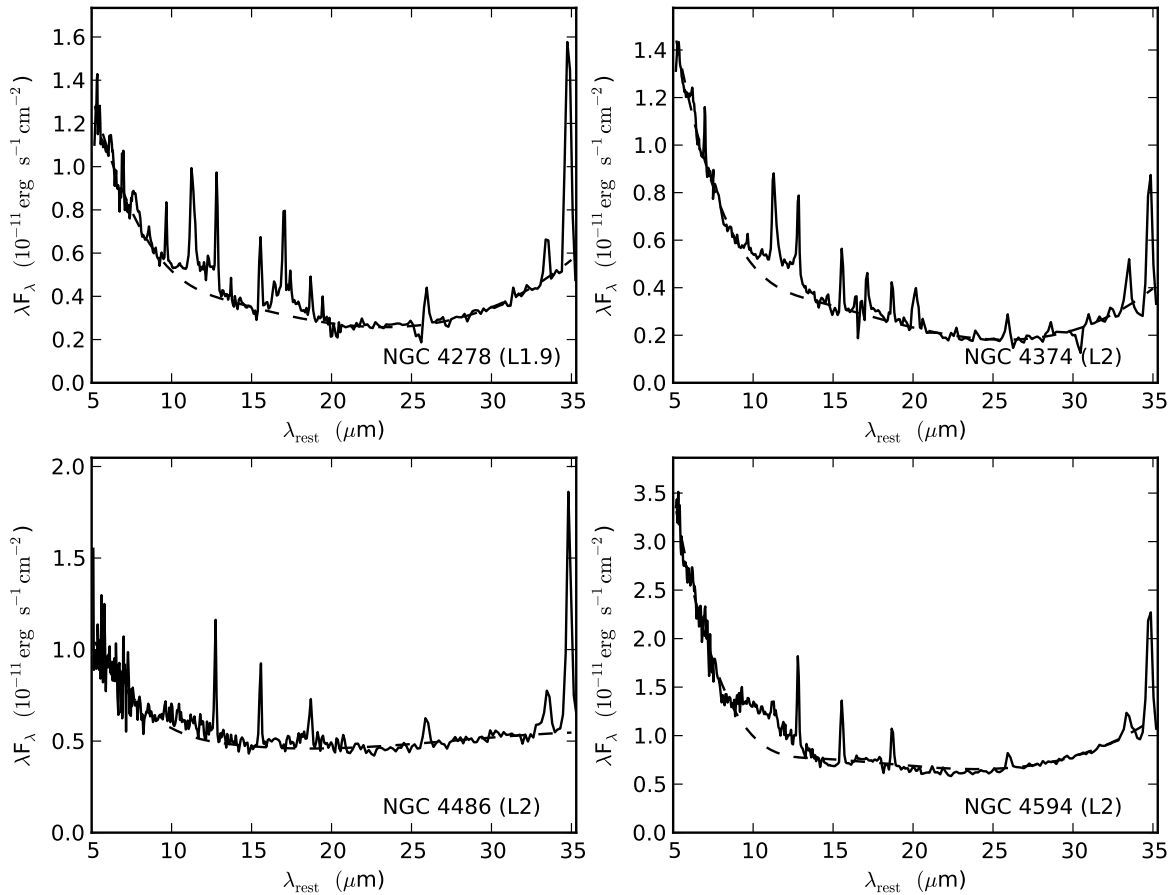}
\caption{ {\small MIR spectra of category II, radio-loud LLAGN with log $\rm L_{bol} / L_{Edd} < -4.6$. Dashed lines show continuum fits used to estimate silicate feature strength.}}
\label{irs2a}
\end{figure*}

\begin{figure*}[htpb]
\includegraphics[scale=0.9]{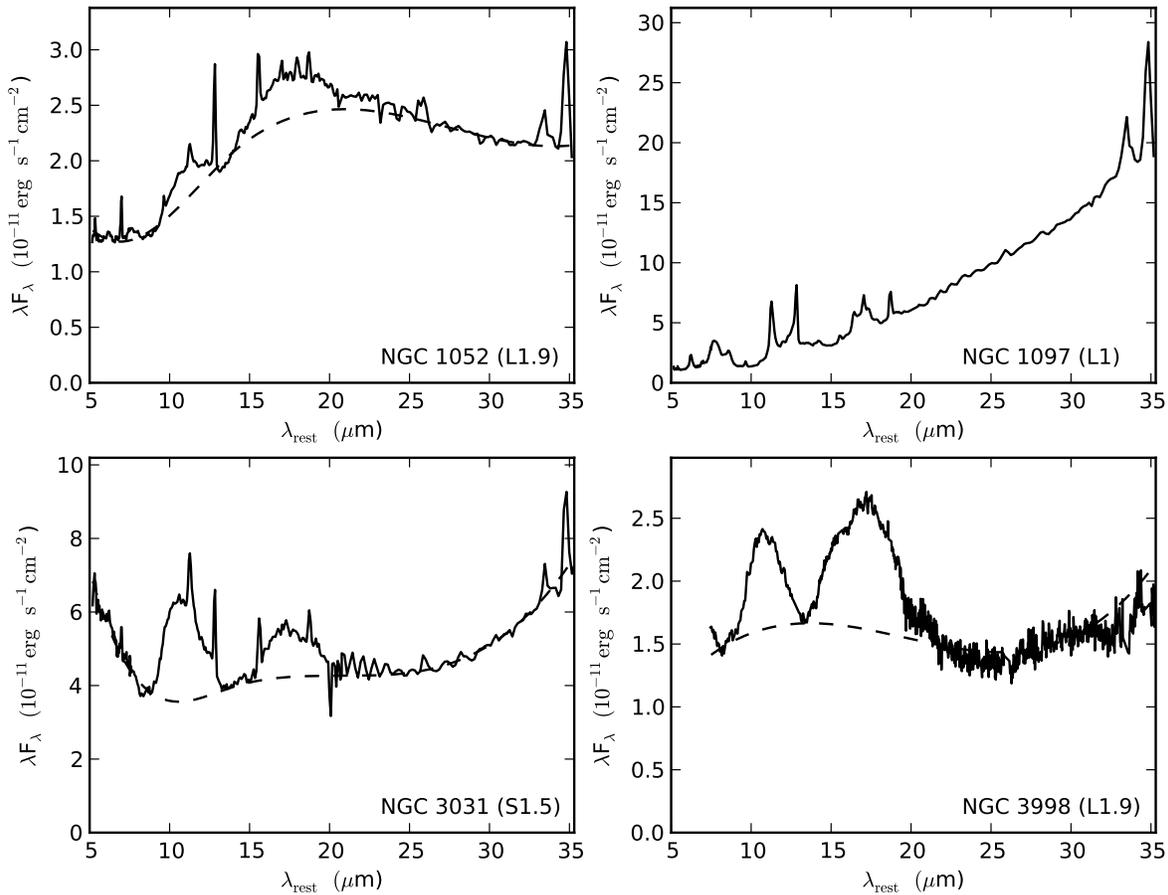}
\caption{ {\small MIR spectra of LLAGN with log $\rm L_{bol} / L_{Edd} > -4.6$ (category III). Dashed lines show continuum fits to objects without strong PAH emission, used to estimate silicate feature strength. The spectrum of NGC~3998 is from \citet{Sturm05} and has had narrow emission features removed.}}
\label{irs3a}
\vspace*{2mm}
\end{figure*}

\begin{figure*}[htpb]
\includegraphics[scale=0.9]{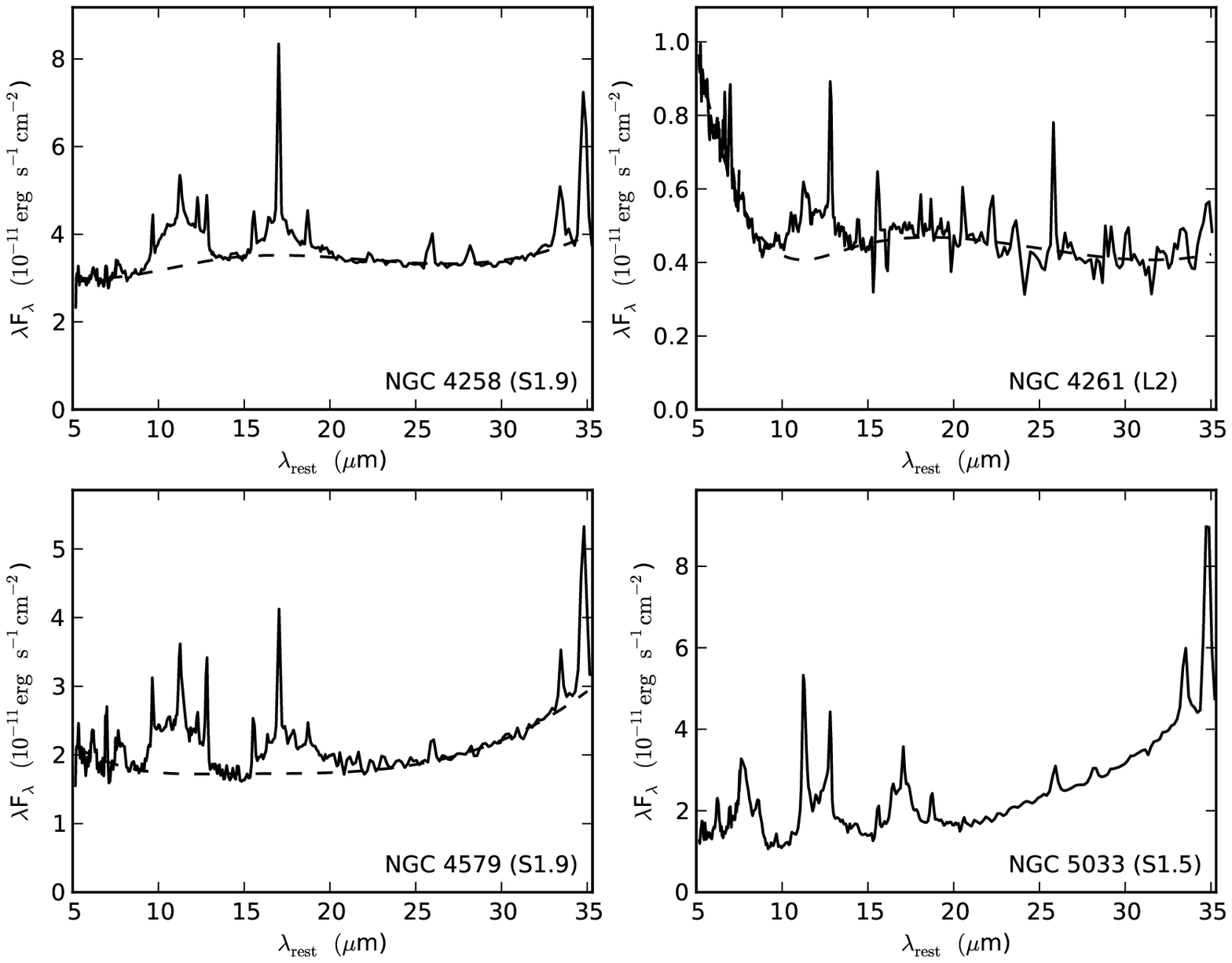}
\caption{ {\small MIR spectra of LLAGN with log $\rm L_{bol} / L_{Edd} > -4.6$ (category III), continued.}}
\label{irs3b}
\end{figure*}

\begin{figure}[htpb]
\includegraphics[scale=0.9, clip, trim=30 200 290 0]{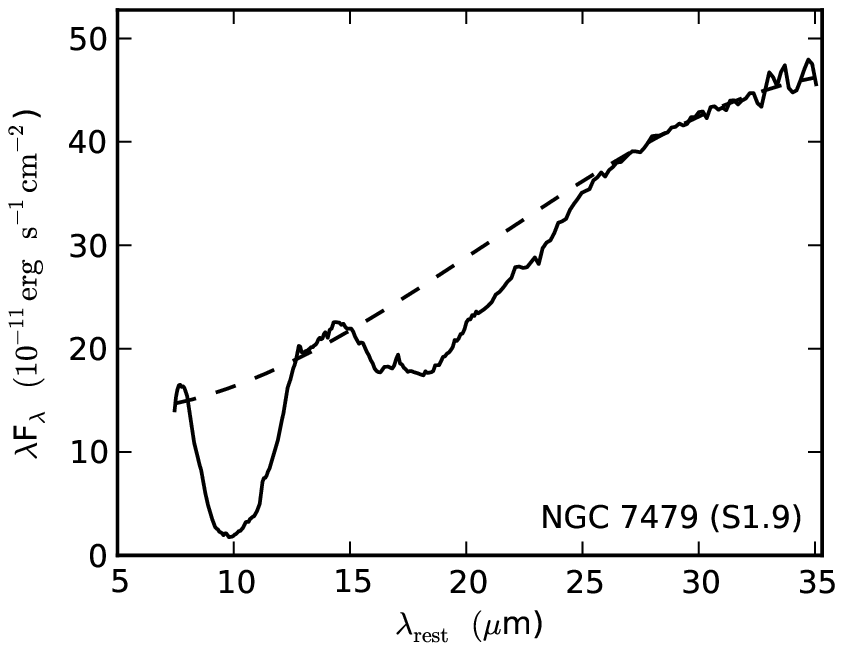}
\caption{ {\small MIR spectra of LLAGN with log $\rm L_{bol} / L_{Edd} > -4.6$ (category III), continued.}}
\label{irs3c}
\end{figure}

The Spitzer/IRS spectra of the LLAGN are shown in Figures \ref{irs1a} -- \ref{irs3c}. The area encompassed by the spectra (3.6\arcsec\ $\times$ 7.2\arcsec\ at 6 $\mu$m) is greater than that shown in Figures \ref{img1} -- \ref{img4}.
It is therefore likely that sources other than the AGN also contribute to the IRS spectra, especially in the faint, diffuse, host-dominated objects.

The ``bowl-shaped'' spectrum of NGC~4594 (Figure \ref{irs2a}) was noted by \citet{Gallimore10}, and the other radio-loud, log $\rm L_{bol} / L_{Edd} < -4.6$ LLAGN have remarkably similar spectra. Silicate emission bands are present in at least three of the spectra, including all three type 2 objects. PAH bands are detected, but are weak in these objects. The five host-dominated, low-$\rm L_{bol} / L_{Edd}$ galaxies with available MIR spectra, on the other hand, all show strong PAH emission and (in those objects with sufficient wavelength coverage) red continua.

Spectra are available for nine of the 12 galaxies with log $\rm L_{bol} / L_{Edd} > -4.6$, and all but one of those galaxies (NGC~4261) have compact morphologies in the ground-based imaging. Of these nine objects, six show prominent silicate emission features. Five of these are type 1.5 -- 1.9 objects, although one (NGC~4261) is a type 2 LINER. Strong PAH emission is observed in a further two objects (NGC~1097 and NGC~5033, both type 1 objects), and the final galaxy in this group, NGC~7479 (S1.5), has very deep silicate absorption bands. NGC~7479 is in fact the only object of the 18 in this sample with IRS spectra in which the silicate features obviously appear in absorption.  The continuum properties of these objects are fairly diverse. The majority of them have somewhat red continua longwards of $\sim 25 \; \mu$m, but the continuum emission of NGC~4261 is essentially flat at $\lambda > 20 \; \mu$m, and that of NGC~1052 is somewhat blue at those wavelengths (in $\lambda F_{\lambda}$).

The strength and frequency of occurrence of the silicate emission features in the sample are remarkable. For most of the high-Eddington ratio and radio-loud, low-Eddington ratio nuclei, we estimate the strength of the 10~$\mu$m silicate feature, $S_{10}$, following \citet{Sirocky08}. The resulting values of $S_{10}$ are reported in Table \ref{measurements}.  \citet{Gallimore10} discuss IRS data reduction and analysis at some length, and present S$_{10}$ for a number of AGN based on a full spectral decomposition. For the two objects in common with their study, we find good agreement in S$_{10}$ (NGC~4579: 0.40 vs 0.40; NGC~4594: 0.31 vs 0.35). Rather than use the simple continuum-fitting method to derive $S_{10}$ for the two log $\rm L_{bol} / L_{Edd} > -4.6$ galaxies with strong PAH emission (NGC~1097 and NGC~5033), we adopt the values given by \citet{Gallimore10} for those objects. Only three of the five host-dominated, low-$\rm L_{bol} / L_{Edd}$ galaxies have full spectral coverage, all have strong PAH bands and the Spitzer spectra of these objects probably contain only a minor contribution from the AGN. For these reasons, we defer measurement of $S_{10}$ in the host-dominated, low-$\rm L_{bol} / L_{Edd}$ galaxies to future work.

\begin{figure}[htpb]
\includegraphics[scale=0.45]{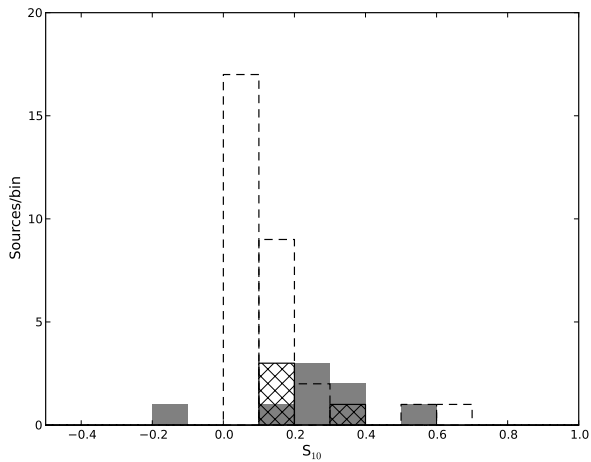}
\caption{ {\small Strength of the 10 $\mu$m silicate emission $S_{10}$ in the high-$\rm L_{bol} / L_{Edd}$  (category III, gray) and radio-loud low-$\rm L_{bol} / L_{Edd}$  (category II, cross-hatched) LLAGN compared with $S_{10}$ in the Seyfert 1 sample of \citet{Thompson09}. For clarity, NGC~7479 from the LLAGN sample and UGC~5101 from the Seyfert 1 sample, which have deep absorption features ($S_{10}$ = -2.19 and -1.52 respectively), are not shown.}}
\label{hist}
\end{figure}

The values of $S_{10}$ in the high- $\rm L_{bol} / L_{Edd}$ nuclei are quite high compared to those commonly observed in type 1 Seyferts of higher luminosity. Figure \ref{hist} compares the silicate emission strengths in the present LLAGN sample and the Seyfert 1-1.5 sample of Thompson et al. (2009; also measured using the method of Sirocky et al. 2008). The LLAGN sample is neither complete nor unbiased, and Figure \ref{hist} is not intended to indicate that the distributions of S$_{10}$ differ systematically between the various AGN types. However, it does demonstrate that the 10~$\mu$m silicate features found in these particular AGN tend to be comparable to the stronger emission features in the Thompson et al. Seyfert 1 sample. 

The very pronounced features in NGC~3031 and NGC~3998 have already drawn some attention in the literature \citep{Sturm05,Smith10}, as has a further strong feature in the type 1 LINER NGC~7213 \citep{Honig10a}. However, strong silicate emission is not known to be a general feature of LLAGN. The 10~$\mu$m emission feature appears to be present in the mean type 1 LINER spectrum of  \citet{Sturm06}, but the authors note that strong silicate emission is not common in that sample, which consists of five objects including two (NGC~3998 and NGC~4278) in common with this paper. As our sample selection was partly based on the existence of previous, large-aperture photometry, objects with strong silicate emission features could in principle have been more likely to be included in this study. As the flux increases by less than a factor of two across the 10~$\mu$m feature, though, it is not clear that this would be a significant effect. In \S\ref{discuss3} we suggest that the strong silicate emission features may represent a particular stage in the evolution of the torus. Analysis of the Spitzer spectra of a larger and/or more targeted sample of LLAGN would be helpful in testing that hypothesis.

\begin{deluxetable*}{ccccccccc}[htpb]
\tablecaption{IR and X-ray properties of the LLAGN. \label{measurements}}
\tablehead{
\colhead{Galaxy} &  F$\rm _{PSF} / F_{2\arcsec}\tablenotemark{a,b}$ & \colhead{$S_{10}$\tablenotemark{c}} & \colhead{$\rm \alpha_{IR}$\tablenotemark{d,b}} & \colhead{log L$_{\rm MIR}$\tablenotemark{b}} &  \colhead{log L$_{\rm 2- 10 keV}$\tablenotemark{e}} & \colhead{$\rm  N_{H}$} & \colhead{Ref.\tablenotemark{f}} & \colhead{CT?\tablenotemark{g}} \\
 & &&  & \colhead{(erg s$^{-1}$)} &\colhead{(erg s$^{-1}$)}  & ($\rm \times 10^{22} cm^{-2}$) & &\\
}
\startdata
NGC~1052 & 0.68 & 0.17 & 2.4 & 41.8 & 41.2 & 12.90 & 1 & N(1)\\
NGC~1097 & 0.93 & 0.21 & 1.4 & 41.4 & 40.6 & 0.023 & 2 & ... \\
NGC~3031 & 0.97 & 0.55 & 0.9 & 40.5 & 40.7 & 0.094 & 3 & N(2) \\
NGC~3166 & 0.43 & ... & ... & 41.2 & 40.7 & ... & 4 & ... \\
NGC~3169 & 0.69 & ... & ... & 41.2 & 41.4 & 11.20 & 5 & ... \\
NGC~3718 & 0.67 & ... & 1.4 & 41.1 & 40.0 & 0.8 & 6& ... \\
NGC~3998 & 0.96 & 0.37 & ... & 41.3 & 41.3 & 2.30 & 1& N(1)\\
NGC~4111 & 0.40 & ... & ... & 40.6 & 39.6 & ... & 7 & N(1)\\
NGC~4258 & 1.00 & 0.30 & 2.5 & 41.1 & 41.0 & 6.8 & 8 & N(2)\\
NGC~4261 & 0.60 & 0.21 & 1.9 & 41.2 & 41.2 & 8.37  & 9 & N(1)\\
NGC~4278 & 0.40 & 0.20 & 1.3 & 39.8 & 40.0 & $<0.035$ & 5 & N(1)\\
NGC~4374 & ...      & 0.19 & ... & $<40.5$ & 39.6 & 0.20 & 7 & Y(1) \\
NGC~4438 & 0.64 & ... & ... & 40.9 & 39.1 & 0.12  & 7 & Y(1) \\
NGC~4457 & 0.50 & ... & ... & 40.9 & 39.0 & 0.1 & 7  & Y(1) \\
NGC~4486 & 0.94 & 0.04 & 1.5 & 41.2 & 40.0 & 0.14  & 10 & N(1) \\
NGC~4579 & 0.88 & 0.40 & ... & 41.5 & 41.2 & 0.93 & 1& N(2) \\
NGC~4594 & 0.25 & 0.31 & ... & 39.9 & 40.0 & 0.19  & 1 & N(1) \\
NGC~4736 & 0.33 & ... & ... & 39.8 & 38.8 & 0.27  & 11 & N(1) \\
NGC~5005 & 0.15 & ... & ... & 40.4 & 40.1 & 3  & 12  & ?(1) \\
NGC~5033 & 0.68 & -0.13 & 0.8 & 41.2 & 41.1 & $<0.03$ & 13 & N(2) \\
NGC~5363 & 0.27 & ... & ... & 40.7 & 39.8 & 2.66 & 1 & Y(1) \\
NGC~7479 & 0.99 & -2.19 & 3.3 & 42.7 & 41.1 & 58 & 14 & ?(2) \\

\enddata
\small{
\tablenotetext{a}{Ratio of MIR PSF-scaling flux to that obtained in a 2\arcsec\ aperture. See \S\ref{mir}.}
\tablenotetext{b}{Based on photometry in the Si-2 (8.8 $\mu$m) filter where available, otherwise N$^{\prime}$ (11.2 $\mu$m), Si-5 (NGC~1097; 11.7 $\mu$m) or N (NGC~4486; 10.8 $\mu$m). See Table \ref{mir-obs} and \S\ref{obsdr}.}
\tablenotetext{c}{Strength of the $\sim10 \;  \mu$m silicate feature (\S\ref{irs}). Positive values of $S_{10}$ indicate emission. Typical errors on $S_{10}$ are estimated to be about 0.05.}
\tablenotetext{d}{NIR -- MIR spectral index, $f_{\nu} \propto \nu^{-\alpha_{IR}}$, for objects possessing at least one MIR and one H band photometric point. See \S\ref{sect:seds}.}
\tablenotetext{e}{Absorption-corrected 2 -- 10 keV luminosity. See \S\ref{sect:seds} for details of X-ray data selection}
\tablenotetext{f}{Reference for X-ray luminosity and column density. 1: \citet{Gonzalez-Martin09a}. 2: \citet{Nemmen06}. 3: \citet{Swartz03}. 4: derived from \citet{Fabbiano92} using standard assumptions to convert from count rate to flux. 5: \citet{Terashima03}. 6: \citet{Satyapal05}. 7: \citet{Flohic06}. 8: \citet{Young04}. 9: \citet{Zezas05}. 10: \citet{Perlman05}. 11: \citet{Eracleous02}. 12: \citet{Guainazzi05}. 13: \citet{Cappi06}. 14: \citet{Panessa06}.}
\tablenotetext{g}{Combination of X-ray and optical indicators suggests galaxy may be Compton-thick. The L$_{X}$ and $N_{H}$ values in columns 5 and 6 have not been modified to account for possible Compton-thickness.   1: \citet{Gonzalez-Martin09b}. 2: \citet{Panessa06}.}
}
\end{deluxetable*}

\begin{deluxetable}{cccc}
\tablecaption{Intrinsic X-ray luminosity estimates for suspected Compton-thick LLAGN. \label{lx_ct}}
\tablehead{ 
\colhead{Galaxy} &  \colhead{$\rm log L_{X}$(spec.)\tablenotemark{a}} & \colhead{$\rm log L_{X}$([O III])\tablenotemark{b}} & \colhead{$\rm log L_{X}$([O IV])\tablenotemark{c} }\\
 &  \colhead{(erg s$^{-1}$)} &  \colhead{(erg s$^{-1}$)} &  \colhead{(erg s$^{-1}$)} \\
}

\startdata

NGC~4374 & 39.6 & 41.3 & 39.3 \\
NGC~4438 & 39.1 & 40.8 & 40.7 \\
NGC~4457 & 39.0 & 40.6 & 40.4 \\
NGC~5005 & 40.1 & 41.6 & 40.6 \\
NGC~5363 & 39.8 & 41.6 & ... \\
NGC~7479 & 41.1 & 42.9 & 41.6 \\

\enddata
\small{
\tablenotetext{a}{ As given in Table \ref{measurements}, from spectral fitting of high-resolution X-ray data. Corrected for absorption, but will nonetheless underestimate the intrinsic X-ray luminosity of a truly Compton-thick AGN.}
\tablenotetext{b}{ ``Corrected'' luminosities from \citet{Gonzalez-Martin09a} and \citet{Panessa06}, based on $\rm F_{X}(2-10 \; keV)/F ([O III])$ ratios}
\tablenotetext{c}{ Based on the $\rm L_{bol, AGN} - L _{[O IV]}$ relation of \citet{Goulding10}, using published [O IV] fluxes \citep{Dudik09,Pereira-Santaella10,Panuzzo11} and assuming L$_{\rm bol} / \rm L_{X} = 16$ \citep[\S\ref{sample}]{Ho09}}
}

\end{deluxetable}

\subsection{Spectral Energy Distributions}
\label{sect:seds}

\subsubsection{Data compilation}

We combine the new IR photometry with published, high-resolution data to produce radio - to - X-ray nuclear SEDs for the LLAGN (Table \ref{seds}; Figures \ref{seds1a} -- \ref{seds3c}). In terms of the nuclear IR emission of LLAGN this SED collection is by far the most complete yet available.

As well as the new MIR data presented here, the SEDs include published, ground-based MIR photometry of several sources. These data were obtained using telescopes of 5 -- 10 m diameter. Ground-based MIR observations are close to diffraction-limited under typical atmospheric conditions, so these measurements generally have resolution $<$0.5\arcsec. All the objects for which we include published MIR data are described as having compact nuclei, and the photometry is either from small apertures \citep[e.g.][]{Horst08} or results from radial profile analysis \citep[e.g.][]{Reunanen10}.

The published 1 -- 5~$\mu$m data are taken from a variety of sources. The shortest-wavelength points are usually from {\em HST} F110W and F160W imaging. The flux from the unresolved component is estimated by fitting a PSF plus exponential, Nuker or other profile to the data \citep[e.g.][]{Quillen01,Ravindranath01,Alonso-Herrero03}, thereby decomposing the image into a stellar and a nuclear component. Emission from stellar photospheres can certainly contribute in the NIR in moderate apertures \citep[e.g.][]{Roche85,Willner85,Clemens11}, but the unresolved flux given by this procedure should represent the AGN alone (unless the nucleus hosts an unresolved stellar cluster). Longwards of 2~$\mu$m, little HST photometry is available and most of the measurements we include are from ground-based imaging. Aside from two galaxies with published adaptive optics imaging \citep[][Fernandez-Ontiveros priv. communication]{Prieto10,Fernandez-Ontiveros11}, the resolution of the ground-based data is generally a few tenths of an arcsecond,  significantly poorer than that of the HST imaging. Nonetheless, an unresolved nuclear flux can be extracted from the data, or an upper limit given \citep[e.g.][]{Alonso-Herrero03}. 

As well as data intended to represent the nuclear emission, lower-resolution photometry has also been added to illustrate the difference between IR SEDs based on large- and small-aperture measurements. These data come from Table \ref{mir-phot}, \citet{Alonso-Herrero01} and \citet{Willner85} and references therein, and use apertures of 2\arcsec, 3\arcsec, and 5 -- 7\arcsec, respectively.

In constructing the non-IR portions of the SEDs we have drawn on the work of \citet{Ho99},  \citet{Nagar05}, \citet{Maoz07} and, especially, \citet{Eracleous10a}, among many others. Additional high-resolution data points have also been included if not already present in the above compilations, and for objects not covered by that work. At all but X-ray frequencies the criterion for inclusion is resolution $<$1\arcsec. 

For the X-ray measurements, we prefer results based on high spatial
resolution data. In particular, Chandra observations (resolution $\sim1\arcsec$)
most effectively spatially isolate the nuclear emission and minimise contamination
from other sources. The effect of pileup in Chandra observations must be considered for bright
sources; for instance, the small-aperture luminosity adopted for NGC~4486 is based on data taken using short frame times designed to avoid pileup problems \citep{Perlman05}. For sources known to be highly variable, we attempt to use observations made
close in time to the MIR observations.  Finally, because the derived X-ray
properties are dependent on the model fitting, where possible we choose from results that fully
describe the applied model and the resulting quality of the fit.

Some of the objects in this sample are suspected to be Compton-thick. In the low-$\rm L_{bol} / L_{Edd}$ categories, \citet{Gonzalez-Martin09b} use several indicators to conclude that NGC~4374, NGC~4438, NGC~4457 and NGC~5363 may have Compton-thick nuclei. High-energy Swift Burst Alert telescope observations of NGC~7479, a high-$\rm L_{bol} / L_{Edd}$ object, suggest an intrinsic 2 -- 10 keV luminosity several times higher than the one adopted in this paper, and modelling of this object's XMM-Newton spectrum implies $N_{H}\sim10^{24} \rm \; cm^{-2}$ \citep{Panessa06,Cusumano10,Brightman11}. Published, extinction-corrected 2 -- 10 keV luminosity estimates for these objects may therefore underestimate their true intrinsic luminosities. Because of the large uncertainties inherent in attempting to correct for possible Compton-thickness, the SED data in Table \ref{seds} and Figures \ref{seds1a} -- \ref{seds3c} have not been adjusted to account for this effect. However, Table \ref{lx_ct} gives possible intrinsic X-ray luminosities for the Compton-thick candidates.

We have also added some larger-aperture ($\sim$7-15\arcsec) submillimeter data to the SEDs. While unlikely to represent emission from the nucleus alone, the data provide a useful consistency check on estimates of jet emission and also give a view of another relatively unexplored spectral region. As well as the SED points themselves, Table \ref{seds} gives the telescope and instrument used for each measurement, with aperture/resolution information where available. The data used in the SEDs were generally obtained over a period of several years, and we note that variability is known to be significant in some of these objects and in some wavelength regimes \citep[e.g.][]{Schoedel07,Pian10}.

\begin{deluxetable*}{ccccccc}[htpb]
\tablecaption{Spectral Energy Distribution Data \label{seds}}
\tablewidth{0pt}
\tablehead{
\colhead{Galaxy} & \colhead{$\nu$} & \colhead{$\nu L_{\nu}$} &  \colhead{Aperture/resolution\tablenotemark{a}} & \colhead{Facility} & \colhead{Reference}  \\ 
& \colhead{(Hz)} & \colhead{(erg s$^{-1}$}) & arcsec &  }
\startdata

NGC 1052 & 5.00E09 & 5.43E+39 & 0.002 & VLBA & Maoz (2007); Kadler et al. (2004) \\
NGC 1052 & 8.40E09 & 9.04E+39 & 0.002 & VLBA & Maoz (2007); Kadler et al. (2004) \\
NGC 1052 & 1.50E10 & 1.41E+40 & 0.0005 & VLBA & Kovalev et al. (2005) \\
NGC 1052 & 2.20E10 & 1.49E+40 & 0.0005 & VLBA & Kadler et al. (2004) \\
NGC 1052 & 4.30E10 & 1.29E+40 & 0.0003 & VLBA & Maoz (2007); Kadler et al. (2004) \\
NGC 1052 & 8.60E10 & 1.07E+40 & 0.001 & VLBI & Lee et al. (2008) \\

\enddata
\tablenotetext{a}{Aperture diameter, resolution or beam size, as described in or inferred from the original paper.}
\tablenotetext{z}{To be published as an online-only table}
\end{deluxetable*}

\subsubsection{Characteristics of the SEDs}
\label{SED_char}

\begin{figure*}[t]
\includegraphics[scale=0.9]{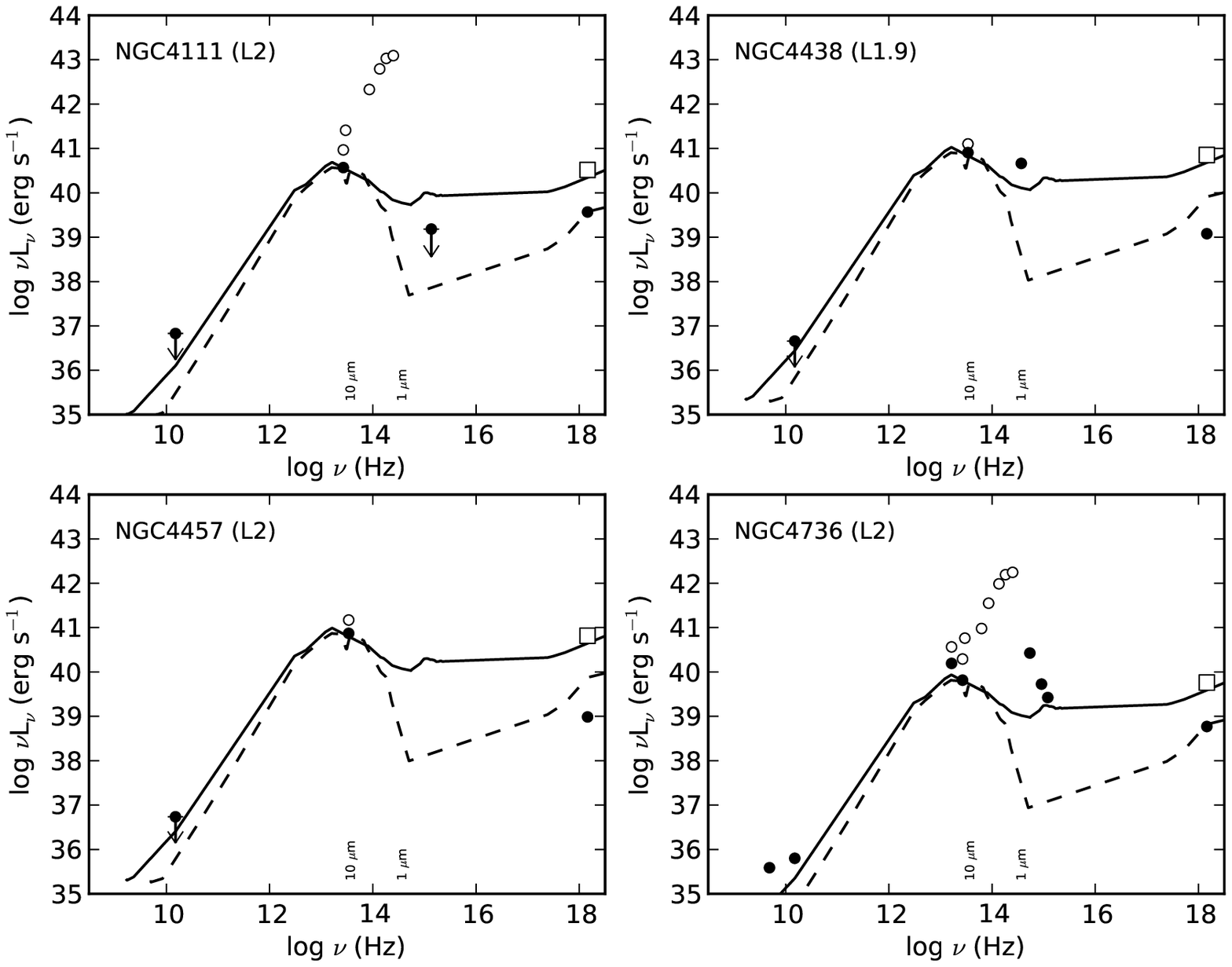}
\caption{ {\small SEDs of host-dominated LLAGN with log $\rm L_{bol} / L_{Edd} < -4.6$ (category I). Open circles denote data with resolution/aperture $>$ 1\arcsec. The X-ray luminosity predicted by the MIR/X-ray relation (\S\ref{mirxsect}) for each galaxy is shown as an open square. Solid and dashed lines mark the mean type 1 and 2 Seyfert SEDs of \citet{Prieto10}, respectively, normalised in the MIR. }}
\label{seds1a}
\end{figure*}

\begin{figure*}[t]
\includegraphics[scale=0.9, clip, trim=0 200 0 0]{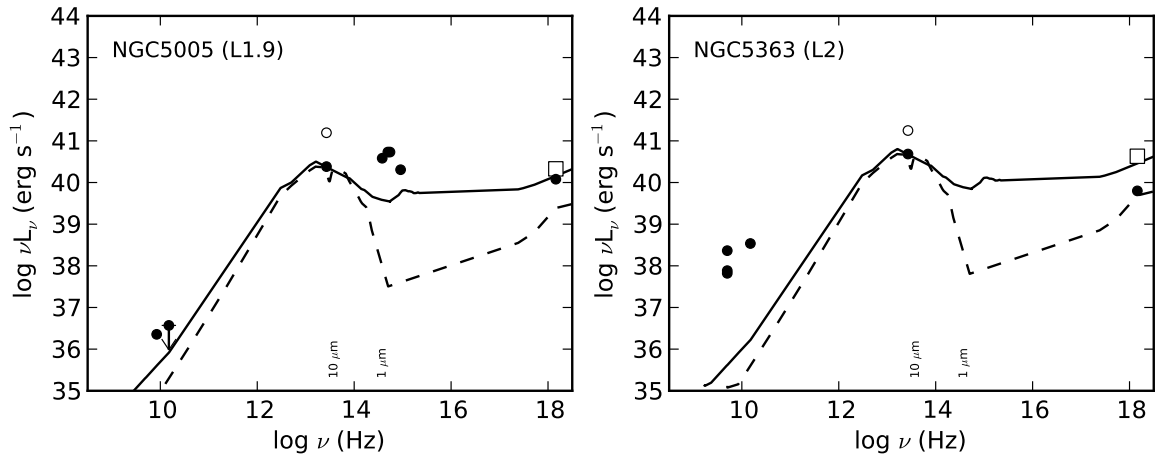}
\caption{ {\small  SEDs of host-dominated LLAGN with log $\rm L_{bol} / L_{Edd} < -4.6$ (category I), continued.}}
\label{seds1b}
\end{figure*}

\begin{figure*}[t]
\includegraphics[scale=0.9]{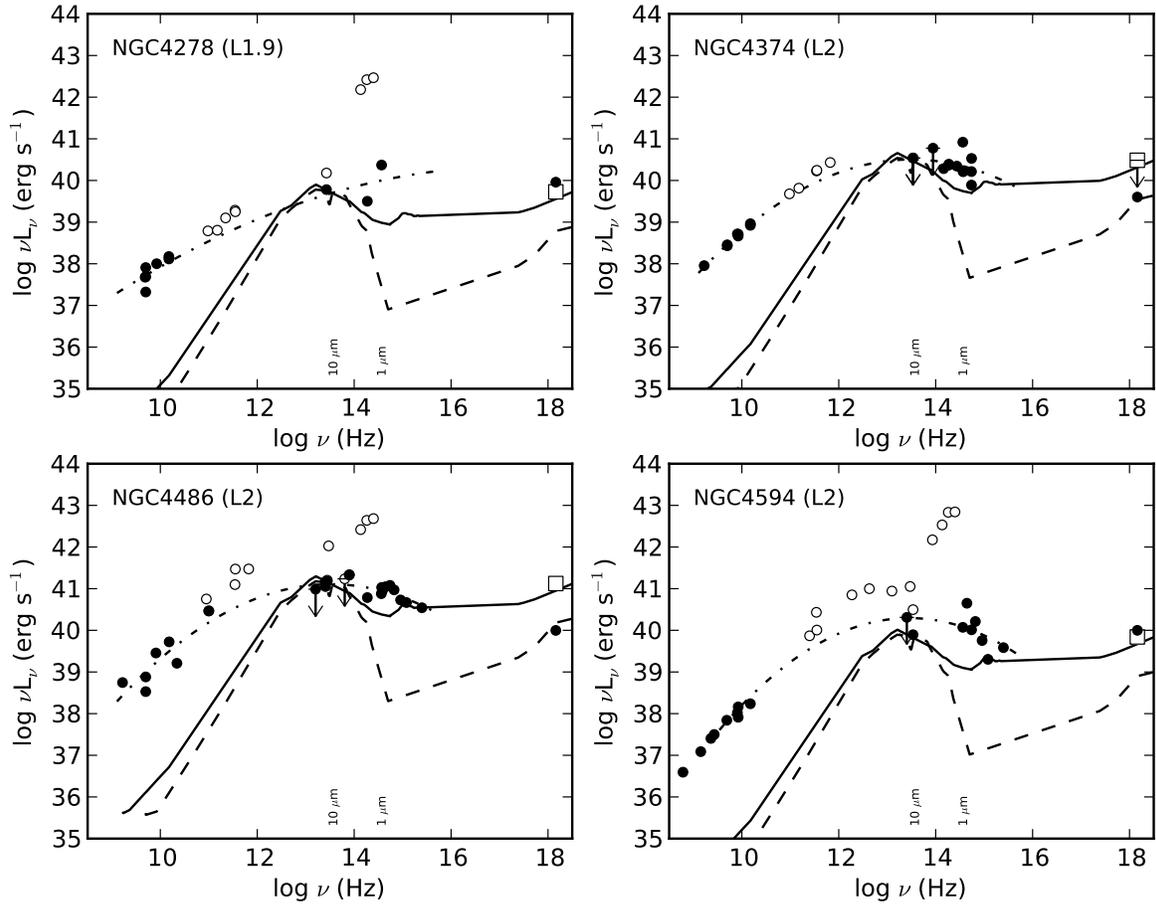}
\caption{ {\small SEDs of radio-loud LLAGN with log $\rm L_{bol} / L_{Edd} < -4.6$ (category II).  Dot-dashed lines show parabolic fits to the nuclear radio/optical/UV data (\S\ref{RLSED}). Other lines and symbols as in Figure \ref{seds1a}.}}
\label{seds2}
\end{figure*}

\begin{figure*}
\includegraphics[scale=0.9]{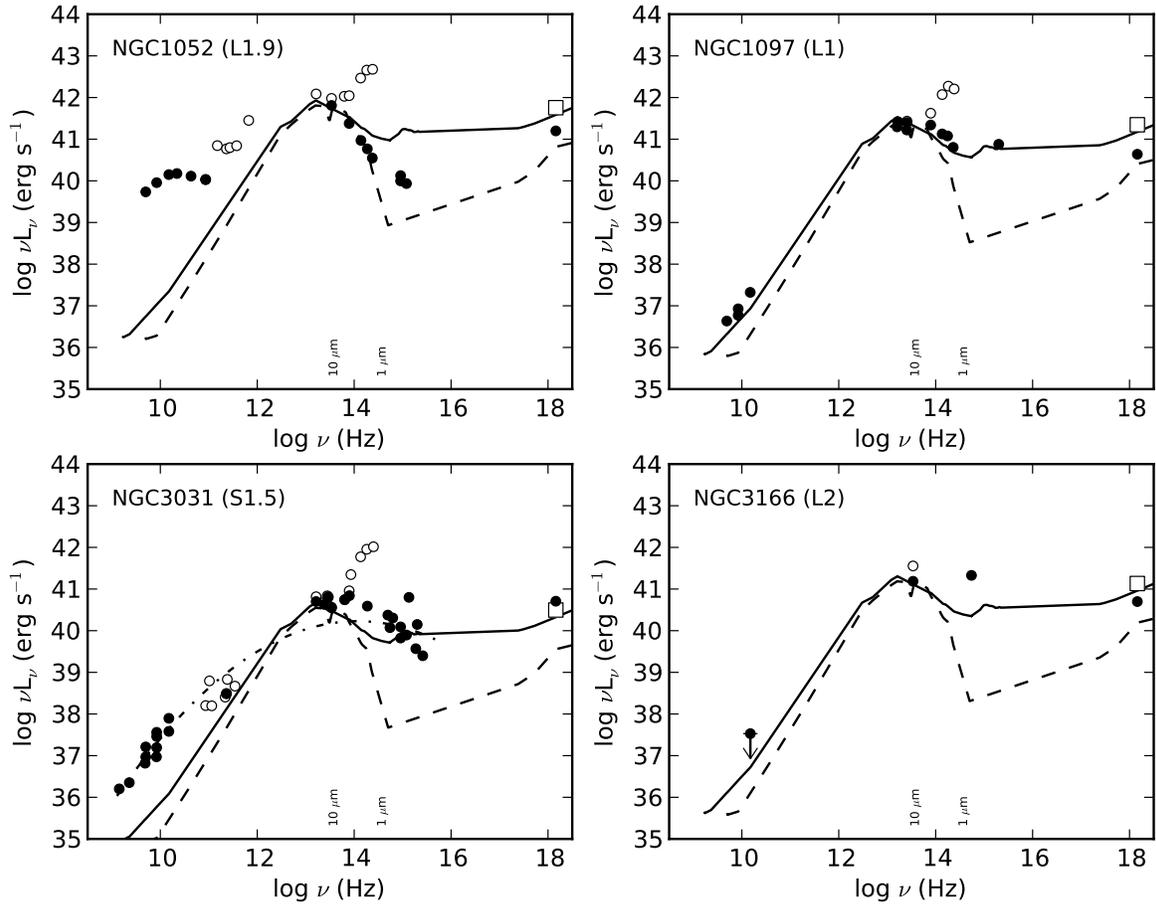}
\caption{ {\small  SEDs of high-Eddington ratio LLAGN (log $\rm L_{bol} / L_{Edd} > -4.6$, category III). Symbols and lines as in Figure \ref{seds2}. }}
\label{seds3a}
\end{figure*}

\begin{figure*}[t]
\includegraphics[scale=0.9]{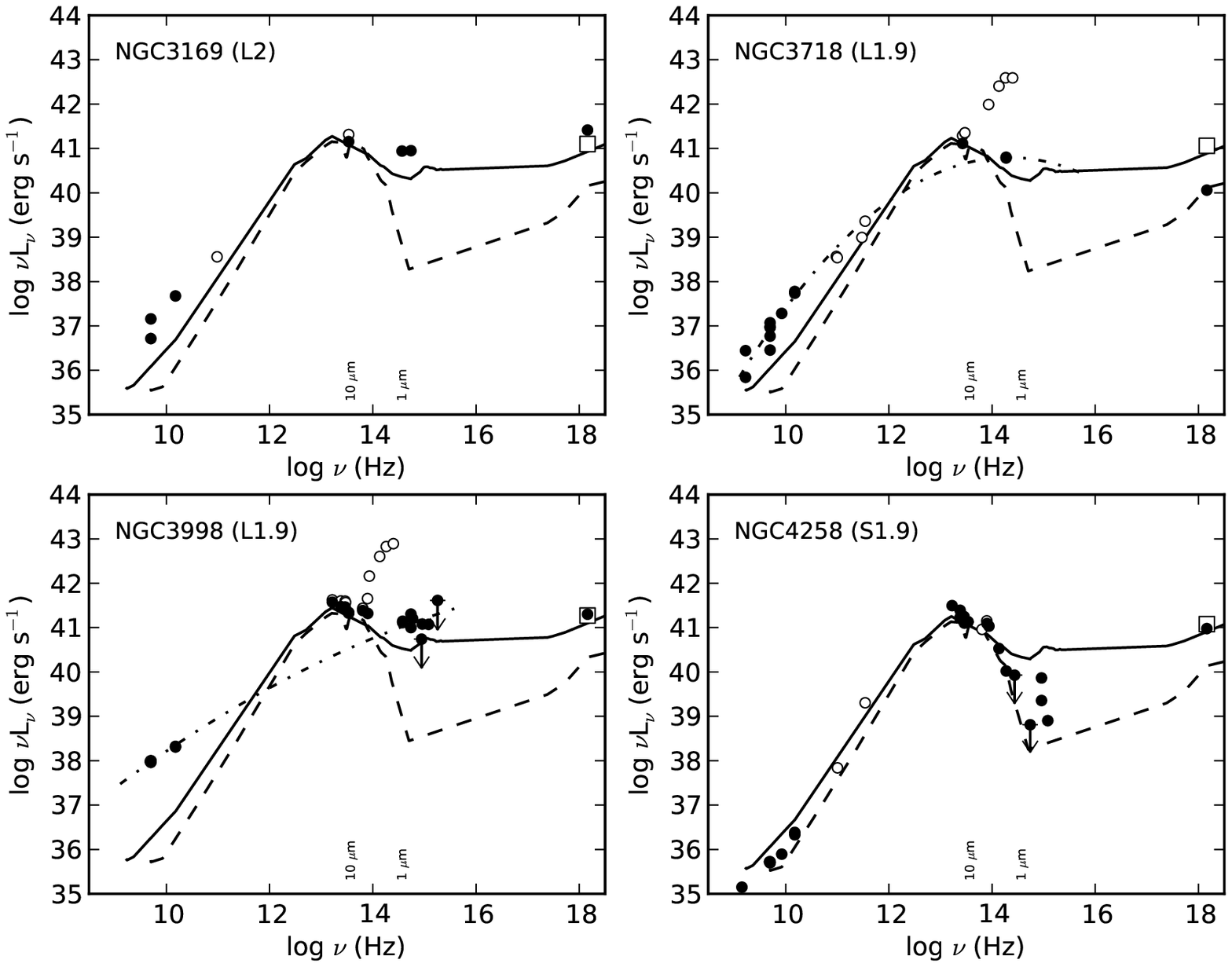}
\caption{ {\small  SEDs of high-Eddington ratio LLAGN (category III), continued.}}
\label{seds3b}
\end{figure*}

\begin{figure}[t]
\includegraphics[scale=0.9]{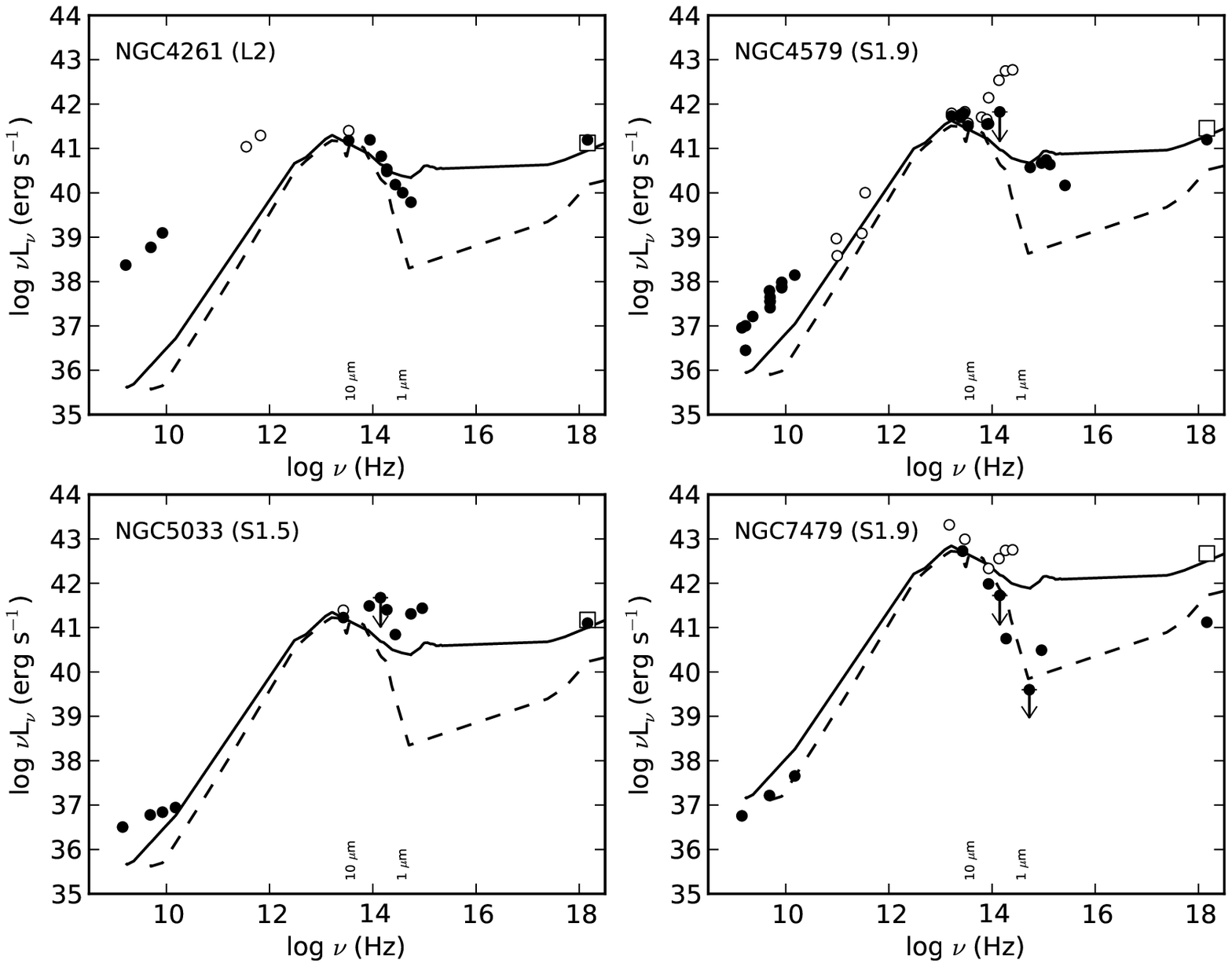}
\caption{ {\small  SEDs of high-Eddington ratio LLAGN (category III), continued.}}
\label{seds3c}
\end{figure}

In Figures \ref{seds1a} -- \ref{seds3c} the SEDs are compared to the mean type 1/2 Seyfert SEDs of \citet{Prieto10}, which result from a careful collection of high-resolution data spanning a wide frequency range. These average SEDs combine data for only 3 -- 4 Seyfert galaxies of each type, so they may not accurately reflect the mean properties of the Seyfert population as a whole. For example, the MIR/X-ray ratios differ between the type 1 and 2 SEDs, contrary to the observation that, in larger samples of Seyferts, this ratio in the various AGN types is statistically indistinguishable \citep{Gandhi09,Levenson09,Honig10a}. Nonetheless, the \citet{Prieto10} Seyfert SEDs serve as a valuable comparison for the overall, multi-wavelength characteristics of the LLAGN SEDs. A close-up view of the 1 -- 20~$\mu$m region for each galaxy possessing at least two IR data points can be found in Appendix \ref{app1}. There, the IR SEDs are compared to the mean 1 -- 20 $\mu$m Seyfert SEDs of \citet{RamosAlmeida09,RamosAlmeida11}.

As expected, the shape of the LLAGN SEDs changes markedly in the IR as the photometric aperture decreases and the measurements are less contaminated by stellar emission. When observed in an aperture of a few arcseconds, almost all of the SEDs are strongly peaked in the NIR. The high-resolution measurements, though, reveal a variety of underlying spectral shapes.

{\bf I. Nuclear SEDs of the host-dominated, low-Eddington ratio LLAGN}

The SEDs of the host-dominated, low-Eddington ratio LLAGN tend to be rather sparsely sampled, with a number of non-detections at radio wavelengths. The available data show a prominent peak in the MIR but, as discussed in \S\ref{obsdr} and  \S\ref{img}, the MIR photometry is unlikely to properly isolate the nucleus in these objects. The objects with high-resolution optical data show rather flat IR-optical spectral slopes. In NGC~4736 and NGC~5005, this may be due to a contribution from a compact stellar cluster in the NIR/optical \citep{Quillen01,MunozMarin07,GonzalezDelgado08}.

{\bf II. Nuclear SEDs of the radio-loud, low-Eddington ratio LLAGN}
\label{RLSED}

The SEDs of these LLAGN are characterised by excess radio emission relative to the average Seyfert galaxy templates, by the lack of a well-defined MIR peak, and by very flat slopes in the MIR -- optical region. Type 2 Seyferts have very steep IR - optical spectra, interpreted as extinction by dust in the torus which remains optically thick well into the IR. As the hot dust in Seyfert 1 nuclei is relatively unobscured, their spectra do not show the same precipitous drop from MIR towards NIR wavelengths \citep{Alonso-Herrero03,Prieto10,RamosAlmeida11}. Despite the fact that 3/4 of the LLAGN in this group are type 2 objects, their MIR - NIR/optical SEDs are at least as flat as that of the mean type 1 Seyfert (Table \ref{measurements}; see also Appendix \ref{app1} for a close-up view of the 1 -- 20 $\mu$m region). 

The four galaxies discussed in this section are among the most radio-loud of the present LLAGN sample, so we may expect a significant contribution from synchrotron radiation to their IR emission. This would have the effect of smearing out any underlying, Seyfert-like MIR peak in the SEDs and could naturally produce a flat MIR - NIR/optical slope. Two of the objects in this category (NGC~4374 and NGC~4486) formed part of the sample used to define the radio-optical correlation of \citet{Chiaberge99}, who interpret the correlation as indicating that the optical luminosity is dominated by synchrotron emission. This suggests that (a) a substantial IR synchrotron contribution is likely, and (b) the core radio and optical SED points may be used to assess the magnitude of that contribution. We therefore use parabolic fits to the radio and optical/UV core data to determine whether the nuclear IR emission is also synchrotron-dominated. This method has been used in a number of studies of radio galaxies  \citep{Cleary07,Leipski09,vanderWolk10}, and \citet{Landau86} show that the radio-to-UV SED of luminous, radio-loud active galaxies can be well fit by such a function. 

As shown in Figure \ref{seds2}, the fits indicate that in all four cases the small-scale IR emission is dominated by the jet. This was already proposed for NGC~4486 (M87) by \citet{Perlman01}. It is also consistent with the results of \citet{vanderWolk10}, who find that most of the FR-I and low-excitation FR-II objects in their sample have MIR emission consistent with a synchrotron spectrum. In the case of NGC~4486 we reiterate that, although the vast majority of the MIR emission is likely to originate in the jet, the expected luminosity of the torus in this object, predicted by the MIR/X-ray relation, is only a small fraction of the observed MIR luminosity (Figure \ref{mirx_all}). The uncertainty in fitting the IR jet contribution is very likely large enough to allow the existence of such a faint torus, so the SED fitting alone does not rule out the presence of a Seyfert-like torus in NGC~4486 and similar radio galaxies. The lack of a steep MIR -- NIR/optical spectral slope is consistent with an unobscured nucleus, but could also be explained by, for instance, significant synchrotron emission on scales larger than a $\sim$pc-scale torus.

On the other hand, the observed MIR emission of NGC~4594 is already consistent with the standard MIR/X-ray relation, and subtracting the estimated synchrotron component would create a significant MIR deficit relative to the expected torus emission in this object. NGC~4594 is therefore likely to be a genuine unobscured type 2 AGN. In all of these objects, however,  even if a weakly-emitting torus does exist, its contribution to the total nuclear MIR luminosity will be negligible for the purpose of even fairly detailed modeling of the broadband SED.

{\bf III. Nuclear SEDs of the high-Eddington ratio LLAGN}

\label{RQSED}

The SEDs of the high-Eddington ratio LLAGN are quite diverse, more so than those of the radio-loud, low-Eddington ratio objects. If they have one distinguishing feature, it is that many of them are more radio-loud than the average Seyfert galaxy. This is expected from previous work showing that radio loudness increases with decreasing Eddington ratio \citep{Ho02,Terashima03,Sikora07}. 

Of the better-sampled SEDs, some -- those of NGC~1052, NGC~4258,  NGC~4261, NGC~4579 and NGC~7479 -- show a well-defined, Seyfert-like peak at MIR wavelengths. The MIR - NIR/optical spectral slopes in these objects are within the range bracketed by the \citet{Prieto10} Seyfert 1 and 2 templates. Amongst these objects, nuclei of the same AGN type do not necessarily have the same spectral slope; for instance, the type 1.9 Seyferts NGC~4258 and NGC~4579. However, if intermediate-type LLAGN have similar nuclei to more luminous intermediate-type objects, then a variety of spectral shapes is to be expected; \citet{Alonso-Herrero03} find a wide range of spectral indices in Seyferts of types 1.8 and 1.9 in the CfA sample. Judged solely by their SEDs, there is little indication that these objects are anything but ``scaled-down'' Seyferts,  complete with a MIR ``dust bump'', albeit with extra radio emission. The high-Eddington ratio objects in this sample are almost exclusively type 1 - 1.9 nuclei, so we cannot search for the systematic differences in SEDs, spectra etc. that are characteristic of type 1 and 2 Seyferts of higher luminosity.

A few of the galaxies -- NGC~1097, NGC~3031, NGC~3718, NGC~3998 and NGC~5033 -- have rather flat MIR -- NIR/optical SEDs compared to the Seyfert templates, and no well-defined MIR peak. These were also characteristics of the radio-loud, low-Eddington ratio objects, whose IR emission in \S\ref{RLSED} we found to be dominated by the jets. In NGC~1097, the ``excess'' NIR/optical emission is probably related to the young, massive nuclear star cluster \citep{StorchiBergmann05,Mason07}. The strong PAH bands and extended 3~$\mu$m emission in NGC~5033 may indicate the presence of nuclear star formation in that object as well \citep{Alonso-Herrero03}. For the remaining three galaxies, we use the fits described in \S\ref{RLSED} to investigate the possible IR synchrotron emission. The fits suggest that, although synchrotron emission is unlikely to dominate the MIR in the same way as for the lower-luminosity, radio-loud, low-Eddington ratio LLAGN, by contributing in the NIR it could flatten the spectral slopes as observed. This fitting procedure assumes that the measured optical luminosities represent optical jet emission, however, which may not be a valid assumption for these objects.

As an alternative means of investigating whether the flat MIR - optical SED slopes are indicative of significant IR synchrotron emission, we searched for a correlation between the IR spectral index (Table \ref{measurements}) and radio loudness (log $\nu L_{\nu} \rm (5 \; GHz)$ / $L_{X}$, \S\ref{sample})
for those galaxies in \S\ref{RLSED} and in this section which have at least one high-resolution NIR data point. We found no relation between these quantities. However, there are several possible ways in which such a correlation could be masked. 
As well as emission from compact clusters such as in NGC~1097, extinction in some objects may mask an intrinsically flat slope. A case in point is NGC~4261, one of the more radio-loud objects in the sample whose nucleus is obscured by $N_{H} \sim 10^{23} \rm cm^{-2}$ \citep{Zezas05} and which is suspected to be heavily extinguished in the optical/UV \citep{Eracleous10a}.  The lack of a correlation could simply reflect that a diversity of phenomena governs the properties of the LLAGN SEDs, and IR observations of a larger sample of objects would be useful in this respect.

To summarise, many of the high-Eddington ratio LLAGN have SEDs that broadly resemble those of conventional Seyferts, although often with enhanced radio emission. In some cases, though, the MIR -- NIR/optical SED slopes are flatter than expected for Seyferts, which hints at possible stellar and/or jet contributions in the IR. Other sources, such as a truncated accretion disk, may also influence the IR properties of the SEDs. Detailed modeling of the SEDs with jet, RIAF, thin disk and torus components is beyond the scope of this paper, but in \S\ref{discuss3} we suggest a scenario which is consistent with the morphological, spectral and broadband SED characteristics of these LLAGN. 

\section {Discussion}
\label{discuss}

We have compiled subarcsecond-resolution, 1 -- 20 $\mu$m imaging of 22 LLAGN, including new IR observations of 20 objects. 
The data have been used to investigate the morphology of the objects and their place on the standard MIR/X-ray plot, and to fill in a region of the SED hitherto lacking high-resolution information. In addition, we have presented Spitzer IR spectroscopy of 18/22 objects.

As a pilot study intending to provide an overview of the IR properties of LLAGN, the sample is somewhat heterogeneous and may be biased towards objects bright in the MIR. Rather than discuss statistical trends within the sample, then, in this section we highlight some interesting properties of the galaxies, loosely divided into three groups with certain common elements.

\subsection{I. The host-dominated, low-Eddington ratio nuclei}
\label{discuss1}

At the low-luminosity end of the sample are nuclei with very low Eddington ratios (log $\rm L_{bol} / L_{Edd} < -4.6$) that are faint compared to the surrounding host galaxy emission. If a dusty torus is present in these objects, it is very likely too faint to contribute significantly to even the high-resolution, PSF-scaling photometric measurements (Figure \ref{detect}). 

All of the host-dominated, low-Eddington ratio galaxies exhibit strong PAH emission in their central regions. Although commonly used to identify starburst galaxies, PAH bands are also observed in many other environments such as the diffuse ISM of dusty elliptical galaxies \citep{Kaneda08}. Three of the five LLAGN in the host-dominated, low-Eddington ratio  category (NGC~4438, NGC~4457 and NGC~5005) are included in the stellar population analysis of \citet{CidFernandes04} and \citet{GonzalezDelgado04}. Intermediate-age stars ($10^{8} - 10^{9}$ yr) contribute $\sim$30 - 60\% of the optical light in these and many of the other LLAGN in those studies, but the contribution of stars aged $< 10^{7}$ yr is $\sim 5\%$ or less. The PAH emission in these objects may be related to the intermediate-age stellar population; based on analysis of the band ratios in a small sample of early-type galaxies, \citet{Vega10} conclude that the PAH molecules are supplied by mass-losing carbon stars formed within the last few Gyr. 
Further detailed and quantitative studies of the PAH bands in LLAGN would be of value in determining whether the features are related to the known intermediate-age population or whether they reveal active star formation that is difficult to detect in conventional optical studies.

\subsection{II. The radio-loud, low-Eddington ratio nuclei}
\label{discuss2}

Also at the low-Eddington ratio, low-luminosity end of the sample, we highlight a set of galaxies with very strong radio emission. The nuclear SEDs of these objects do not show the prominent mid-IR peak observed in ``conventional'' Seyfert galaxies, and their MIIR-NIR/optical spectral slopes are flatter than that of the average type 1 Seyfert. We find that the nuclear IR emission is dominated by synchrotron radiation from the jet, consistent with results for various radio galaxy samples \citep{Leipski09,vanderWolk10}. 

High-resolution SEDs can show whether the IR emission is energetically dominated by nonthermal processes. However, additional information is needed to conclusively rule out the presence of a Seyfert-like torus. The well-known MIR/X-ray relation can be used to estimate the MIR emission that would be expected from the torus, and in at least one galaxy (NGC~4594) the observed MIR luminosity is already no more than would be expected from a torus, before accounting for the likely dominant MIR synchrotron component. This galaxy is therefore probably an example of a genuine, ``bare'' type 2 AGN with no obscuring torus \citep[also suggested by the low X-ray column, $2 \times 10^{21} \rm \; cm^{-2}$;][]{Gonzalez-Martin09a}. The same may be true for NGC~4486. However, the emission from a torus heated by the weak AGN in NGC~4486 would account for only $\sim$10\% of the observed luminosity, well within the error of the SED fitting used to investigate the synchrotron component. In terms of the IR emission there is therefore room to ``hide'' a torus in NGC~4486. Additional evidence, such as the minimal X-ray absorption towards this type 2 nucleus \citep{DiMatteo03,Perlman05} is needed to support any assertion that NGC~4486 is an unobscured type 2 AGN.

The IR spectra of the galaxies in this category, 3/4 of which are type 2 objects, all appear to show silicate emission. The PAH and [Ne II] emission in NGC~4278 mean that the presence of the silicate features in this object is ambiguous, and the 11.3~$\mu$m PAH band in NGC~4374 could contribute significantly to the measured value of S$_{10}$ in that nucleus. However, the PAH features in NGC~4486 and NGC~4594 are weak and the silicate emission in NGC~4486 is well-known \citep{Perlman07,Buson09}. Silicate dust features are an unambiguous sign that dust is present. In higher-luminosity AGN, silicate emission features are characteristic of type 1 objects and are an expected result of a direct view of hot dust in a roughly face-on torus \citep{Nenkova08b}. Silicate emission is also known in a few type 2 AGN and can be reproduced by clumpy torus models \citep{Mason09,Nikutta09,Alonso-Herrero11}. This explanation seems unlikely for the radio-loud, low-Eddington ratio nuclei for two reasons. First, silicate emission is rare in type 2 Seyferts \citep[e.g.][]{Shi06}, so it would be surprising to encounter it in all three type 2 objects in this sample. Second, as noted above, some of these galaxies probably do not host a torus.

If not arising in a standard, Seyfert-like torus, the silicate emission bands could arise in diffuse, optically thin dust perhaps associated with the remains of a dissipating torus. We develop this line of reasoning in \S\ref{discuss3}. Alternatively, the features may be produced in circumstellar dust shells. The weak, extended silicate emission in several early-type galaxies has been shown to originate in mass-losing stars \citep{Bressan06}, and this process is also likely to explain the silicate features in NGC~4486 \citep{Buson09}.

\subsection{III. The high-Eddington ratio nuclei}
\label{discuss3}

The nuclei with log $\rm L_{bol} / L_{Edd} > -4.6$, which lie at the high luminosity end of the sample, tend to have prominent nuclear point sources in the MIR. They often have strong silicate emission features compared to those typically observed in Seyfert 1 nuclei, and their MIR luminosities are consistent with or slightly in excess of those predicted by the standard MIR/X-ray relation. The nuclear broadband SEDs of these objects are rather mixed. Some are essentially indistinguishable from those of ``conventional'' Seyferts, many have excess radio emission compared to higher-luminosity Seyferts, and a few have unusually flat MIR - NIR/optical slopes. 

\begin{figure}[h]
\includegraphics[scale=0.45]{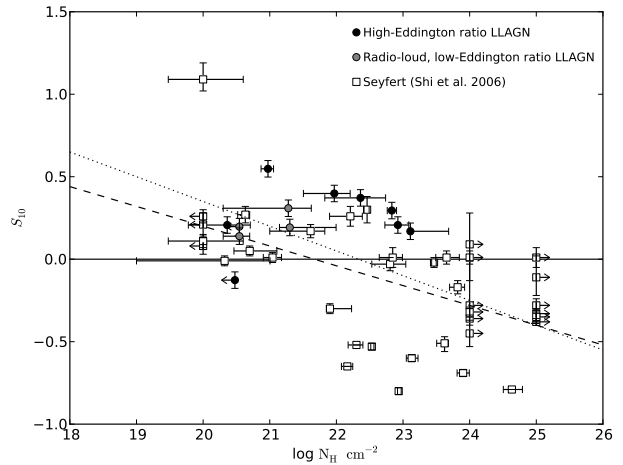}
\caption{ {\small Strength of the 10 $\mu$m silicate feature vs X-ray column density for the high-Eddington ratio (category III) and radio-loud, low-Eddington ratio (category II) LLAGN. Square symbols indicate the Seyfert galaxies of \citet{Shi06}. The dashed line shows Shi et al.'s fits to their Seyferts, the dotted line their fit to their whole sample (mostly comprised of Seyferts and various classes of quasar). Positive values of $S_{10}$ indicate emission, negative values absorption. As in Figure \ref{hist}, NGC~7479, which has a deep absorption feature ($S_{10} = -2.19$), is omitted for clarity.}}
\label{NH}
\end{figure}

Considered in isolation, these characteristics offer no compelling evidence that the torus is absent in these LLAGN or that it differs from that observed in higher-luminosity Seyferts. However, we do find some indications that the IR emission may not arise in the standard, Seyfert-like torus of the AGN unified model. In Figure \ref{NH}, we plot the strength of the 10~$\mu$m silicate feature against HI column density, for both the Seyfert galaxies of \citet{Shi06} and the LLAGN with S$_{10}$ measurements (\S\ref{irs}, Table \ref{measurements}). In the Shi et al. Seyferts, S$_{10}$ and N$_{\rm H}$ are loosely correlated. The scatter presumably reflects the fact that the observed $S_{10}$ is likely a complicated function of the precise arrangement of clouds in the torus \citep{Honig10b}, and may also indicate a contribution from absorption in the host galaxy \citep{Roche07,Deo09,Alonso-Herrero11}. Many of the high-Eddington ratio LLAGN lie on the upper envelope of the Seyfert points, with relatively high values of $S_{10}$ per unit N$_{H}$.  One possible explanation is that the dust-to-gas ratio is lower in these particular objects than in most Seyfert galaxies. This may be expected if the torus is an optically thick region of an accretion disk wind that becomes less powerful at low accretion rates \citep{Elitzur09}, when the amount of material reaching the dust sublimation radius and able to form dust grains is reduced. In this case we may expect either a torus with fewer clouds and a higher probability of observing a hot, directly-illuminated cloud face, or simply optically thin dust emission. Both of these configurations would cause relatively strong silicate emission features, consistent with our finding that S$_{10}$ in many of these high-Eddington ratio LLAGN is strong compared to that in typical type 1 Seyferts (Figure \ref{hist}).

As mentioned in \S\ref{discuss2}, a  contribution to the observed silicate features from circumstellar shells in the host galaxy is also possible. As the luminosity of the central engine diminishes, the relative strength of features arising in the surrounding stellar population will increase. This could also account for changes in $S_{10}$/N$_{\rm H}$ among LLAGN.

The high-Eddington ratio LLAGN have at least as much MIR emission as predicted by the MIR/X-ray relation for Seyferts and quasars. If the torus in these LLAGN does indeed contain less dust, the IR continuum emission must be produced by some other mechanism. A contribution from synchrotron radiation is likely, and may explain the unusually flat MIR -NIR/optical slopes observed in some of the SEDs. Another possibility is that some of the IR emission comes from a truncated accretion disk. In the models of \citet{Nemmen11}, the disk emission peaks between 1 -- 10~$\mu$m and can account for essentially all the luminosity at these wavelengths. Emission associated with nuclear star clusters, such as that known to exist in NGC~1097 \citep{StorchiBergmann05} may also play a r\^{o}le. However, \citet{Asmus11} find that star formation contributes $<$30\% of the small-scale 12~$\mu$m flux in most LLAGN of comparable luminosity to those studied in this paper. Detailed modelling of the LLAGN, taking advantage of the new constraints from the well-sampled IR SEDs presented here, will provide more insight into the processes responsible for the IR emission of LLAGN and the role of the torus in these objects.

\section{Conclusions}

The new, high-resolution IR imaging and SEDs presented in this paper represent the first step towards establishing the nuclear IR properties of a significant number of LLAGN. At the median distance of the galaxies studied here, 16.8 Mpc, the AGN is well isolated in objects with log L$_{2-10 \rm \; keV} \gtrsim 40.5 \rm \; erg \; s^{-1}$ ($\rm log \; L_{bol} \gtrsim 41.8 \rm \; erg \; s^{-1}$). Deeper observations would allow a cleaner detection of the pointlike central engine in less luminous or more distant objects, but the cost in observing time would be large with current facilities.

The Spitzer spectra of the less radio-loud of the lowest-luminosity LLAGN show strong PAH features. Detailed analysis of the band ratios etc. may demonstrate a connection with very young stars or active star formation, rather than the well-known intermediate-age stellar population or the diffuse ISM. If so, combining MIR and optical spectroscopy of a larger sample could provide a more complete view of the star formation histories of LLAGN.

Some of the lower-luminosity, low-Eddington ratio galaxies are very strong radio emitters. The high-resolution SEDs of these objects lack a well-defined MIR peak and are very flat in the MIR -- NIR/optical range. The broadband SEDs allow us to conclude that their IR emission is dominated by synchrotron radiation from jets. Some of these objects, such as NGC~4594 (the Sombrero galaxy) are probably unobscured type 2 objects, genuinely lacking a broad line region. In other objects, such as NGC 4486 (M87), the torus emission predicted by the standard AGN MIR/X-ray relation is so weak that a torus cannot be ruled out based on SED information alone; supporting information is also needed. 

At relatively high Eddington ratios and luminosities, the galaxies have prominent, compact MIR nuclei. Their SEDs are quite diverse, ranging from objects indistinguishable from typical type 2 Seyferts to nuclei with strong radio emission and flat MIR - NIR/optical slopes. The strong silicate emission features present in many of these objects demonstrate that dust is present in their nuclear regions. However, we tentatively suggest that those strong features and the values of S$_{10}$/N$_{H}$ in these nuclei may indicate optically thin dust and low dust-to-gas ratios, consistent with models predicting the disappearance of the torus in LLAGN. If so, much of the IR continuum emission must be caused by processes other than emission by dust in the torus. Possible sources include synchrotron radiation, a truncated accretion disk and nuclear star formation. Observationally, IR polarimetry and monitoring campaigns may constrain the r\^{o}le of each of these processes.  A targeted study of the silicate features in a well-defined LLAGN sample should also help clarify the properties of the torus in these objects. Finally, we expect that detailed modelling of the data in terms of accretion disk, RIAF, jets and torus will be valuable in elucidating the nature of the IR emission of LLAGN.

\acknowledgments
We are grateful to the anonymous referee for a clear and helpful report, and we would also like to thank J. A. Fern\'{a}ndez-Ontiveros for providing data in advance of publication. REM appreciates the hospitality of the University of Sheffield during a portion of this work. This paper is based on observations obtained at the Gemini Observatory, which is operated by the
Association of Universities for Research in Astronomy, Inc., under a cooperative agreement
with the NSF on behalf of the Gemini partnership: the National Science Foundation (United
States), the Science and Technology Facilities Council (United Kingdom), the
National Research Council (Canada), CONICYT (Chile), the Australian Research Council
(Australia), MinistŽrio da Cincia e Tecnologia (Brazil) and SECYT (Argentina). A.A.-H. and L. C. acknowledge support from the Spanish
Plan Nacional de Astronom\'{\i}a y Astrof\'{\i}sica under grants
AYA2009-05705-E and AYA2010-21161-C02-1. CRA acknowledges the Spanish Ministry of Science and Innovation (MICINN) through project 
Consolider-Ingenio 2010 Program grant CSD2006-00070: First Science with the GTC 
(http://www.iac.es/consolider-ingenio-gtc/) and PN AYA2010-21887-C04.04

\appendix

\section{Appendix 1: The 1 -- 20~$\mu$m region of the SEDs}
\label{app1}

\begin{figure}[t]
\includegraphics[scale=0.9, clip, trim=0 0 0 0]{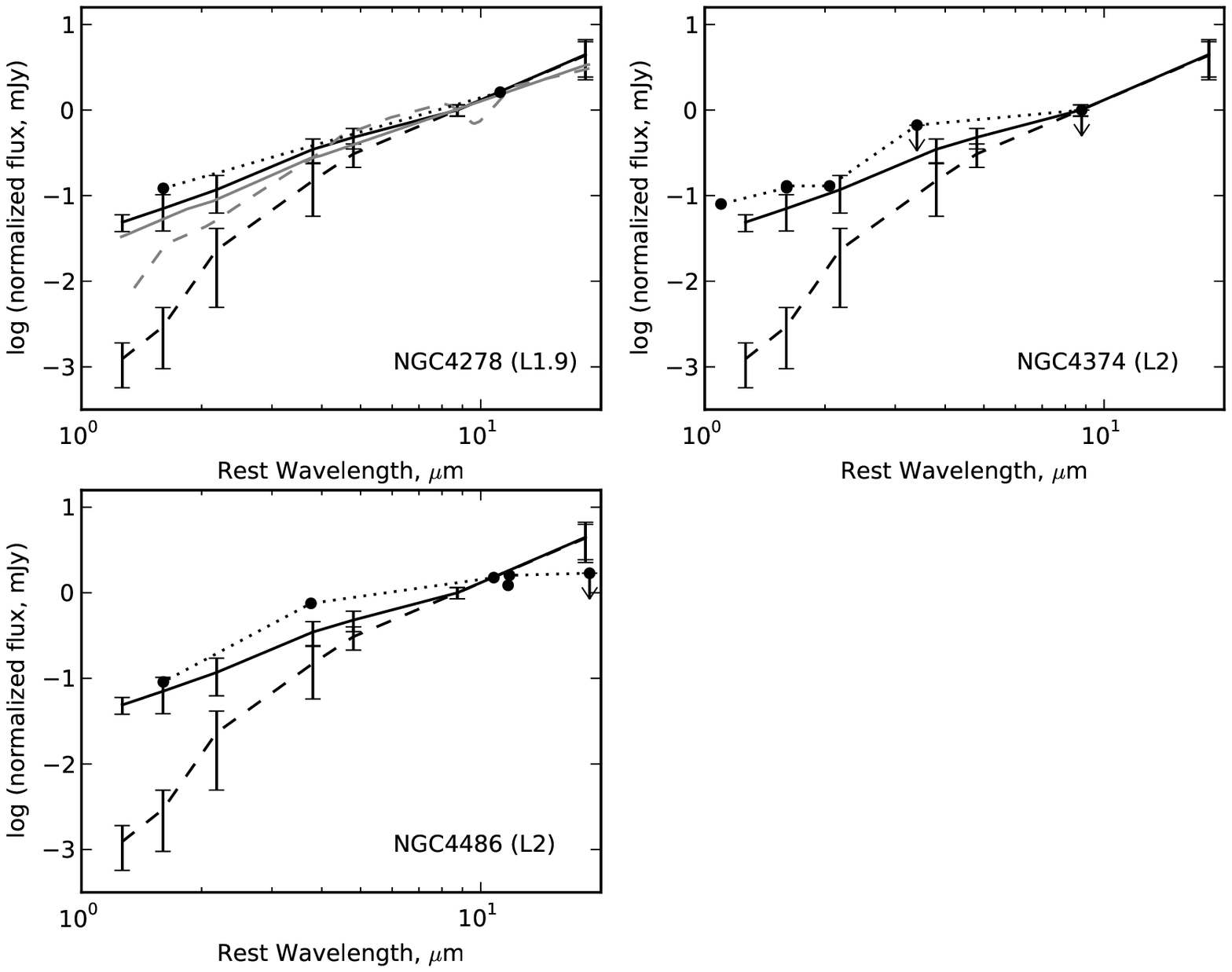}
\caption{{\small 1 -- 20 $\mu$m SEDs of the radio-loud, low-Eddington ratio (category II) LLAGN having at least two IR photometric points. The solid black line shows the mean type 1 Seyfert SED of \citet{RamosAlmeida11}, the dashed black line the mean type 2 Seyfert SED of \citet{RamosAlmeida09}. The LLAGN are denoted by the dotted line and filled circle symbols. For NGC~4278 only, the average Seyfert 1 and 2 SEDs of \citet{Prieto10} are plotted in gray. The template SEDs are normalised at 8.7 $\mu$m. The observed SEDs are normalised either directly to the Si2 ($\approx$ 8.7 $\mu$m) photometric point, if available, otherwise using the filter closest in wavelength and assuming $\rm \alpha_{MIR}=2.0$ \citep{RamosAlmeida11}. Errors on the observed data are not shown as error estimates are not available for many of the photometric points from the literature. Where available, errors are typically comparable to the size of the plot symbols.}}
\label{zoom1a}
\end{figure}

The high-resolution IR data are the novel aspect of this paper, so in this Appendix we present figures focusing on the 1 -- 20~$\mu$m part of the SEDs. Rather than the mean Seyfert SEDs of \citet{Prieto10} used in \S\ref{sect:seds}, in this section the LLAGN SEDs are compared to the mean SEDs of \citet{RamosAlmeida09,RamosAlmeida11}. The \citet{Prieto10} templates have the advantage of wider wavelength coverage, but those of \citet{RamosAlmeida09,RamosAlmeida11} include more objects. We therefore judge that the Ramos Almeida data are more suitable for a detailed comparison of the 1 -- 20~$\mu$m region. Both sets of SEDs are shown in Figure \ref{zoom1a}. While the type 1 SEDs agree fairly well (and would be indistinguishable on the scale used in Figures \ref{seds1a} -- \ref{seds3c}), the type 2 SED of \citet{Prieto10} is significantly less steep than that of \citet{RamosAlmeida09}. The difference in the type 2 IR SEDs does not affect the discussion in \S\ref{sect:seds}, which covers a broad wavelength range and highlights differences between some of the LLAGN and the average type 1 Seyfert. IR spectral indices for these LLAGN are given in Table \ref{measurements}. For comparison, \citet{RamosAlmeida11} give mean values of $1.6 \pm 0.2$ for type 1 Seyferts and  $3.6 \pm 0.8$ for type 2 Seyferts.

\begin{figure}
\includegraphics[scale=0.9]{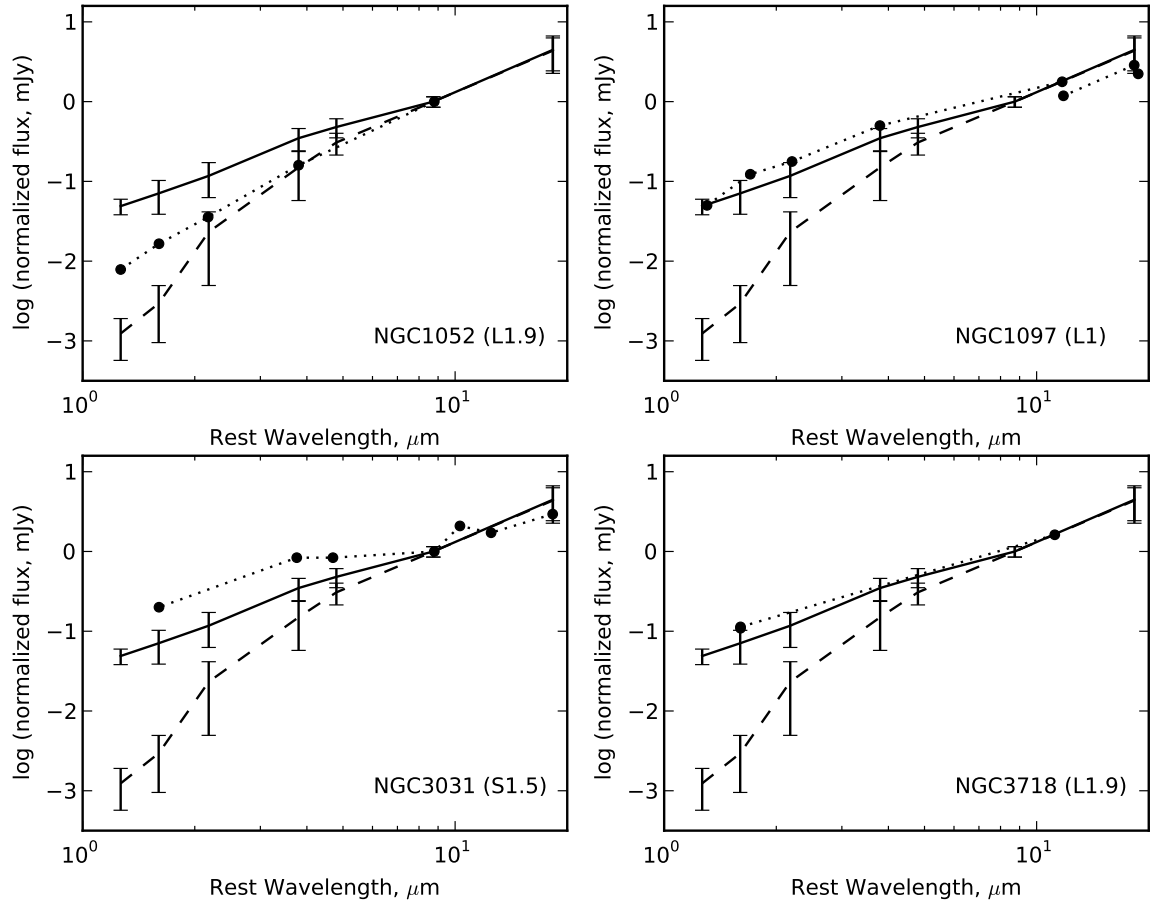}
\caption{ {\small  1 -- 20 $\mu$m SEDs of the high-Eddington ratio (category III) LLAGN having at least two IR photometric points. Lines and symbols as in Figure \ref{zoom1a}.}}
\label{zoom2a}
\end{figure}

\begin{figure}[t]
\includegraphics[scale=0.9]{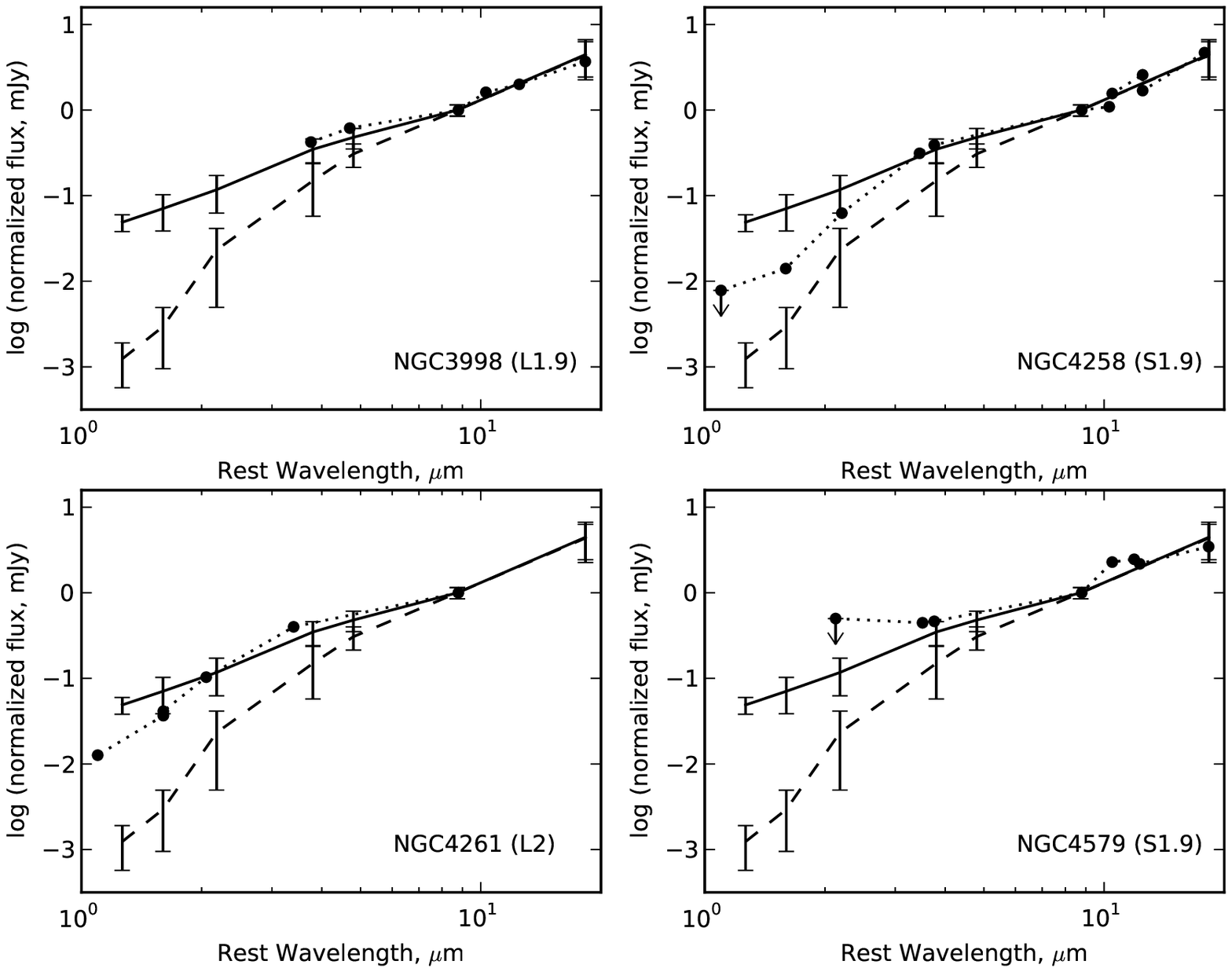}
\caption{ {\small  1 -- 20 $\mu$m SEDs of the high-Eddington ratio (category III) LLAGN having at least two IR photometric points, continued.}}
\label{zoom2b}
\end{figure}

\begin{figure}[t]
\includegraphics[scale=0.9, clip, trim=0 190 50 30]{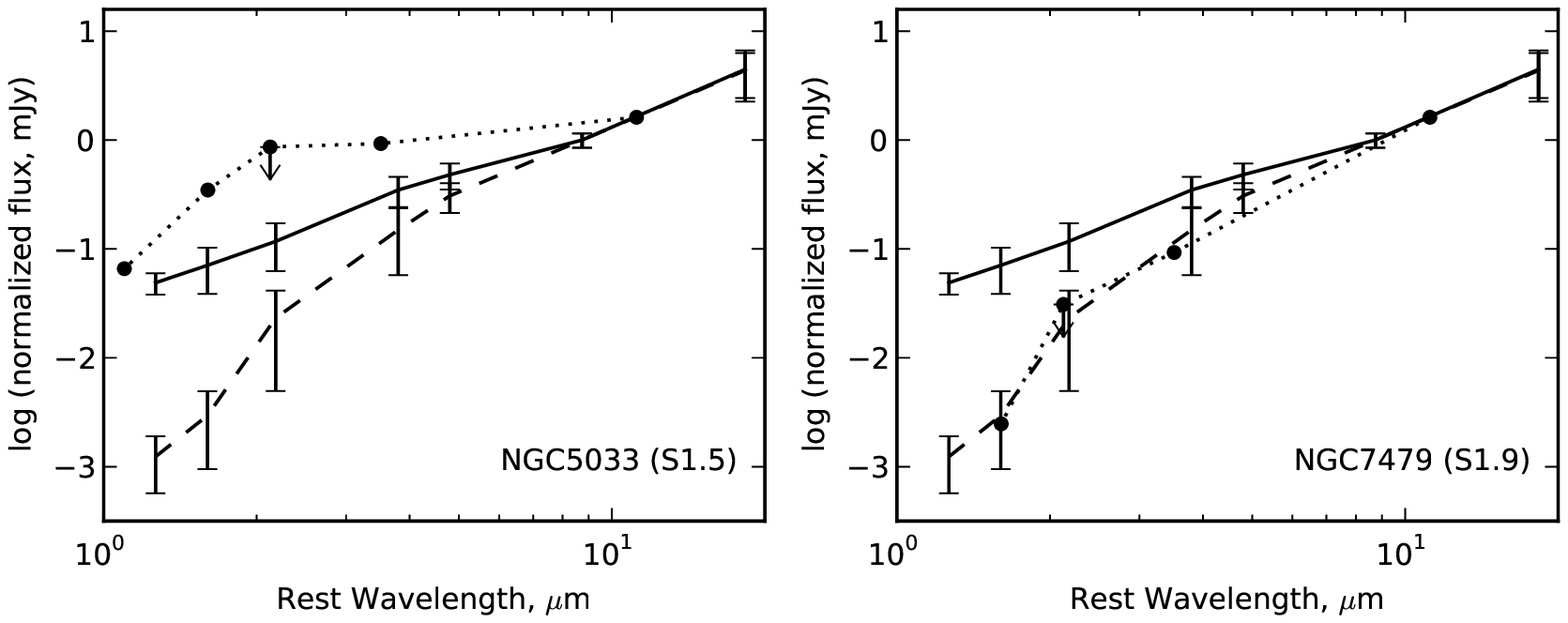}
\caption{ {\small  1 -- 20 $\mu$m SEDs of the high-Eddington ratio (category III) LLAGN having at least two IR photometric points, continued.}}
\label{zoom2c}
\end{figure}


\end{document}